\documentclass{article}

\usepackage{breakurl}

\usepackage{amsmath,amsthm,amsfonts,amssymb}
\usepackage{fontenc}
\usepackage[all]{xy}
\usepackage{graphicx,subfigure,multirow}
\usepackage[figurename=Fig.]{caption}
\usepackage{tikz}
\usetikzlibrary{shapes, arrows, shadows}
\usepackage[latin1]{inputenc}
\usepackage[scaled]{helvet}
\usepackage[toc, page]{appendix}
\usepackage[colorinlistoftodos]{todonotes}

\usepackage{times}
\usepackage{latexsym}
\usepackage{dsfont}
\usepackage{yfonts}
\usepackage{tikz}
\usepackage[normalem]{ulem}

\usepackage[a4paper]{geometry}
\usepackage[colorlinks=true, linkcolor=blue]{hyperref}

\usetikzlibrary{circuits.ee.IEC}
\usetikzlibrary{decorations.pathmorphing}
\usetikzlibrary{shapes.geometric}

\tikzstyle{env}=[copoint,regular polygon rotate=0,minimum width=0.2cm, fill=black]

\tikzstyle{probs}=[shape=semicircle,fill=white,draw=black,shape border rotate=180,minimum width=1.2cm]

\usetikzlibrary{circuits.ee.IEC}
\usetikzlibrary{decorations.pathmorphing}
\usetikzlibrary{shapes.geometric}

\tikzstyle{wavy}=[decorate,decoration={snake, segment length=1mm, amplitude=0.3mm}]
\tikzstyle{mopoint}=[shape=semicircle, fill=white,draw=black,shape border rotate=180,minimum width = 0.9cm,scale =0.75, xscale=0.7]
\tikzstyle{mocopoint}=[shape=semicircle, fill=white,draw=black,minimum width = 0.9cm, scale =0.75, xscale=0.7]

\tikzstyle{cpoint}=[shape=semicircle, fill=white,draw=black,minimum width = 0.9cm, scale =0.75, xscale=0.7, shape border rotate = 90]
\tikzstyle{cocpoint}=[shape=semicircle, fill=white,draw=black,minimum width = 0.9cm, scale =0.75, xscale=0.7, shape border rotate = 90]

\tikzstyle{every picture}=[baseline=-0.25em,scale=0.5]
\tikzstyle{dotpic}=[] 
\tikzstyle{diredges}=[every to/.style={diredge}]
\tikzstyle{math matrix}=[matrix of math nodes,left delimiter=(,right delimiter=),inner sep=2pt,column sep=1em,row sep=0.5em,nodes={inner sep=0pt},text height=1.5ex, text depth=0.25ex]

\tikzstyle{inline text}=[text height=1.5ex, text depth=0.25ex,yshift=0.5mm]
\tikzstyle{label}=[font=\footnotesize,text height=1.5ex, text depth=0.25ex,yshift=0.5mm]
\tikzstyle{left label}=[label,anchor=east,xshift=1.5mm]
\tikzstyle{right label}=[label,anchor=west,xshift=-1.5mm]

\tikzstyle{braceedge}=[decorate,decoration={brace,amplitude=2mm,raise=-1mm}]
\tikzstyle{small braceedge}=[decorate,decoration={brace,amplitude=1mm,raise=-1mm}]

\tikzstyle{doubled}=[line width=2pt] 
\tikzstyle{boldedge}=[doubled,shorten <=-0.17mm,shorten >=-0.17mm]

\tikzstyle{boldedgedashed}=[very thick,dashed,shorten <=-0.17mm,shorten >=-0.17mm]
\tikzstyle{vboldedgedashed}=[doubled,dashed,shorten <=-0.17mm,shorten >=-0.17mm]
\tikzstyle{left hook arrow}=[left hook-latex]
\tikzstyle{right hook arrow}=[right hook-latex]
\tikzstyle{sembracket}=[line width=0.5pt,shorten <=-0.07mm,shorten >=-0.07mm]

\tikzstyle{causal edge}=[->,thick,gray]
\tikzstyle{causal nondir}=[thick,gray]
\tikzstyle{timeline}=[thick,gray, dashed]

\tikzstyle{cedge}=[<->,thick,gray!70!white]

\tikzstyle{empty diagram}=[draw=gray!90!white,dashed,shape=rectangle,minimum width=1cm,minimum height=1cm]
\tikzstyle{empty diagram small}=[draw=gray!50!white,dashed,shape=rectangle,minimum width=0.6cm,minimum height=0.5cm]

\tikzstyle{dot}=[inner sep=0.7mm,minimum width=0pt,minimum height=0pt,draw,shape=circle]
\tikzstyle{ddot}=[inner sep=0.7mm,doubled, minimum width=2.5mm,minimum height=2.5mm,draw,shape=circle]

\tikzstyle{black dot}=[dot,fill=black]
\tikzstyle{white dot}=[dot,fill=white]
\tikzstyle{green dot}=[white dot] 
\tikzstyle{gray dot}=[dot,fill=gray!40!white]
\tikzstyle{red dot}=[gray dot] 

\tikzstyle{black ddot}=[ddot,fill=black]
\tikzstyle{white ddot}=[ddot,fill=white]
\tikzstyle{gray ddot}=[ddot,fill=gray!40!white]

\tikzstyle{gray edge}=[gray!40!white]

\tikzstyle{small dot}=[inner sep=0.4mm,minimum width=0pt,minimum height=0pt,draw,shape=circle]

\tikzstyle{small black dot}=[small dot,fill=black]
\tikzstyle{small white dot}=[small dot,fill=white]
\tikzstyle{small gray dot}=[small dot,fill=gray!40!white]

\tikzstyle{causal dot}=[inner sep=0.4mm,minimum width=0pt,minimum height=0pt,draw=white,shape=circle,fill=gray!40!white]

\tikzstyle{white phase dot}=[dot,fill=white]
\tikzstyle{white phase ddot}=[ddot,fill=white]

\tikzstyle{gray phase dot}=[dot,fill=gray!40!white]
\tikzstyle{gray phase ddot}=[ddot,fill=gray!40!white]
\tikzstyle{grey phase dot}=[gray phase dot]
\tikzstyle{grey phase ddot}=[gray phase ddot]

\tikzstyle{cnot}=[fill=white,shape=circle,inner sep=-1.4pt]
\tikzstyle{hadamard}=[square box,inner sep=0 pt,font=\tiny\sf,minimum height=3mm,minimum width=3mm]
\tikzstyle{dhadamard}=[hadamard,doubled]
\tikzstyle{antipode}=[white dot,inner sep=0.3mm,font=\footnotesize]

\tikzstyle{scalar}=[diamond,draw,inner sep=0.5pt,font=\small]
\tikzstyle{dscalar}=[diamond,doubled, draw,inner sep=0.5pt,font=\small]

\tikzstyle{small box}=[rectangle,inline text,fill=white,draw,minimum height=5mm,yshift=-0.5mm,minimum width=5mm,font=\small]
\tikzstyle{small gray box}=[small box,fill=gray!30]
\tikzstyle{medium box}=[rectangle,inline text,fill=white,draw,minimum height=5mm,yshift=-0.5mm,minimum width=10mm,font=\small]
\tikzstyle{square box}=[small box] 
\tikzstyle{medium gray box}=[small box,fill=gray!30]
\tikzstyle{large box}=[rectangle,inline text,fill=white,draw,minimum height=5mm,yshift=-0.5mm,minimum width=15mm,font=\small]
\tikzstyle{large gray box}=[small box,fill=gray!30]

\tikzstyle{point}=[regular polygon,regular polygon sides=3,draw,scale=0.75,inner sep=-0.5pt,minimum width=9mm,fill=white,regular polygon rotate=180]
\tikzstyle{copoint}=[regular polygon,regular polygon sides=3,draw,scale=0.75,inner sep=-0.5pt,minimum width=9mm,fill=white]
\tikzstyle{dpoint}=[point,doubled]
\tikzstyle{dcopoint}=[copoint,doubled]

\tikzstyle{tinypoint}=[regular polygon,regular polygon sides=3,draw,scale=0.55,inner sep=-0.15pt,minimum width=6mm,fill=white,regular polygon rotate=180]

\tikzstyle{white point}=[point]
\tikzstyle{green point}=[white point] 
\tikzstyle{white copoint}=[copoint]
\tikzstyle{gray point}=[point,fill=gray!40!white]
\tikzstyle{gray dpoint}=[gray point,doubled]
\tikzstyle{red point}=[gray point] 
\tikzstyle{gray copoint}=[copoint,fill=gray!40!white]
\tikzstyle{gray dcopoint}=[gray copoint,doubled]

\tikzstyle{tiny gray point}=[tinypoint,fill=gray!40!white]

\tikzstyle{diredge}=[->]
\tikzstyle{rdiredge}=[<-]
\tikzstyle{thickdiredge}=[->, very thick]
\tikzstyle{pointer edge}=[->,very thick,gray]
\tikzstyle{pointer edge part}=[very thick,gray]
\tikzstyle{dashed edge}=[dashed]
\tikzstyle{thick dashed edge}=[very thick,dashed]
\tikzstyle{thick gray dashed edge}=[thick dashed edge,gray!40]
\tikzstyle{thick map edge}=[very thick,|->]

\makeatletter
\newcommand{\boxshape}[3]{%
\pgfdeclareshape{#1}{
\inheritsavedanchors[from=rectangle] 
\inheritanchorborder[from=rectangle]
\inheritanchor[from=rectangle]{center}
\inheritanchor[from=rectangle]{north}
\inheritanchor[from=rectangle]{south}
\inheritanchor[from=rectangle]{west}
\inheritanchor[from=rectangle]{east}
\backgroundpath{
\southwest \pgf@xa=\pgf@x \pgf@ya=\pgf@y
\northeast \pgf@xb=\pgf@x \pgf@yb=\pgf@y

\@tempdima=#2
\@tempdimb=#3

\pgfpathmoveto{\pgfpoint{\pgf@xa - 5pt + \@tempdima}{\pgf@ya}}
\pgfpathlineto{\pgfpoint{\pgf@xa - 5pt - \@tempdima}{\pgf@yb}}
\pgfpathlineto{\pgfpoint{\pgf@xb + 5pt + \@tempdimb}{\pgf@yb}}
\pgfpathlineto{\pgfpoint{\pgf@xb + 5pt - \@tempdimb}{\pgf@ya}}
\pgfpathlineto{\pgfpoint{\pgf@xa - 5pt + \@tempdima}{\pgf@ya}}
\pgfpathclose
}
}}

\boxshape{NEbox}{0pt}{5pt}
\boxshape{SEbox}{0pt}{-5pt}
\boxshape{NWbox}{5pt}{0pt}
\boxshape{SWbox}{-5pt}{0pt}
\boxshape{EBox}{-3pt}{3pt}
\boxshape{WBox}{3pt}{-3pt}
\makeatother

\tikzstyle{cloud}=[shape=cloud,draw,minimum width=1.5cm,minimum height=1.5cm]

\tikzstyle{map}=[draw,shape=NEbox,inner sep=2pt,minimum height=6mm,fill=white]
\tikzstyle{mapdag}=[draw,shape=SEbox,inner sep=2pt,minimum height=6mm,fill=white]
\tikzstyle{mapadj}=[draw,shape=SEbox,inner sep=2pt,minimum height=6mm,fill=white]
\tikzstyle{maptrans}=[draw,shape=SWbox,inner sep=2pt,minimum height=6mm,fill=white]
\tikzstyle{mapconj}=[draw,shape=NWbox,inner sep=2pt,minimum height=6mm,fill=white]

\tikzstyle{dbox}=[draw,doubled,shape=rectangle,inner sep=2pt,minimum height=6mm,minimum width=6mm,fill=white]
\tikzstyle{dmap}=[draw,doubled,shape=NEbox,inner sep=2pt,minimum height=6mm,fill=white]
\tikzstyle{dmapdag}=[draw,doubled,shape=SEbox,inner sep=2pt,minimum height=6mm,fill=white]
\tikzstyle{dmapadj}=[draw,doubled,shape=SEbox,inner sep=2pt,minimum height=6mm,fill=white]
\tikzstyle{dmaptrans}=[draw,doubled,shape=SWbox,inner sep=2pt,minimum height=6mm,fill=white]
\tikzstyle{dmapconj}=[draw,doubled,shape=NWbox,inner sep=2pt,minimum height=6mm,fill=white]

\tikzstyle{ddmap}=[draw,doubled,dashed,shape=NEbox,inner sep=2pt,minimum height=6mm,fill=white]
\tikzstyle{ddmapdag}=[draw,doubled,dashed,shape=SEbox,inner sep=2pt,minimum height=6mm,fill=white]
\tikzstyle{ddmapadj}=[draw,doubled,dashed,shape=SEbox,inner sep=2pt,minimum height=6mm,fill=white]
\tikzstyle{ddmaptrans}=[draw,doubled,dashed,shape=SWbox,inner sep=2pt,minimum height=6mm,fill=white]
\tikzstyle{ddmapconj}=[draw,doubled,dashed,shape=NWbox,inner sep=2pt,minimum height=6mm,fill=white]

\boxshape{sNEbox}{0pt}{3pt}
\boxshape{sSEbox}{0pt}{-3pt}
\boxshape{sNWbox}{3pt}{0pt}
\boxshape{sSWbox}{-3pt}{0pt}
\tikzstyle{smap}=[draw,shape=sNEbox,fill=white]
\tikzstyle{smapdag}=[draw,shape=sSEbox,fill=white]
\tikzstyle{smapadj}=[draw,shape=sSEbox,fill=white]
\tikzstyle{smaptrans}=[draw,shape=sSWbox,fill=white]
\tikzstyle{smapconj}=[draw,shape=sNWbox,fill=white]

\tikzstyle{dsmap}=[draw,dashed,shape=sNEbox,fill=white]
\tikzstyle{dsmapdag}=[draw,dashed,shape=sSEbox,fill=white]
\tikzstyle{dsmaptrans}=[draw,dashed,shape=sSWbox,fill=white]
\tikzstyle{dsmapconj}=[draw,dashed,shape=sNWbox,fill=white]

\boxshape{mNEbox}{0pt}{10pt}
\boxshape{mSEbox}{0pt}{-10pt}
\boxshape{mNWbox}{10pt}{0pt}
\boxshape{mSWbox}{-10pt}{0pt}
\tikzstyle{mmap}=[draw,shape=mNEbox]
\tikzstyle{mmapdag}=[draw,shape=mSEbox]
\tikzstyle{mmaptrans}=[draw,shape=mSWbox]
\tikzstyle{mmapconj}=[draw,shape=mNWbox]

\tikzstyle{mmapgray}=[draw,fill=gray!40!white,shape=mNEbox]
\tikzstyle{smapgray}=[draw,fill=gray!40!white,shape=sNEbox]

\makeatletter
\pgfdeclareshape{cornerpoint}{
\inheritsavedanchors[from=rectangle] 
\inheritanchorborder[from=rectangle]
\inheritanchor[from=rectangle]{center}
\inheritanchor[from=rectangle]{north}
\inheritanchor[from=rectangle]{south}
\inheritanchor[from=rectangle]{west}
\inheritanchor[from=rectangle]{east}
\backgroundpath{
\southwest \pgf@xa=\pgf@x \pgf@ya=\pgf@y
\northeast \pgf@xb=\pgf@x \pgf@yb=\pgf@y

\pgfmathsetmacro{\pgf@shorten@left}{\pgfkeysvalueof{/tikz/shorten left}}
\pgfmathsetmacro{\pgf@shorten@right}{\pgfkeysvalueof{/tikz/shorten right}}

\pgfpathmoveto{\pgfpoint{0.5 * (\pgf@xa + \pgf@xb)}{\pgf@ya - 5pt}}
\pgfpathlineto{\pgfpoint{\pgf@xa - 8pt + \pgf@shorten@left}{\pgf@yb - 1.5 * \pgf@shorten@left}}
\pgfpathlineto{\pgfpoint{\pgf@xa - 8pt + \pgf@shorten@left}{\pgf@yb}}
\pgfpathlineto{\pgfpoint{\pgf@xb + 8pt - \pgf@shorten@right}{\pgf@yb}}
\pgfpathlineto{\pgfpoint{\pgf@xb + 8pt - \pgf@shorten@right}{\pgf@yb - 1.5 * \pgf@shorten@right}}
\pgfpathclose
}
}

\pgfdeclareshape{cornercopoint}{
\inheritsavedanchors[from=rectangle] 
\inheritanchorborder[from=rectangle]
\inheritanchor[from=rectangle]{center}
\inheritanchor[from=rectangle]{north}
\inheritanchor[from=rectangle]{south}
\inheritanchor[from=rectangle]{west}
\inheritanchor[from=rectangle]{east}
\backgroundpath{
\southwest \pgf@xa=\pgf@x \pgf@ya=\pgf@y
\northeast \pgf@xb=\pgf@x \pgf@yb=\pgf@y

\pgfmathsetmacro{\pgf@shorten@left}{\pgfkeysvalueof{/tikz/shorten left}}
\pgfmathsetmacro{\pgf@shorten@right}{\pgfkeysvalueof{/tikz/shorten right}}

\pgfpathmoveto{\pgfpoint{0.5 * (\pgf@xa + \pgf@xb)}{\pgf@yb + 5pt}}
\pgfpathlineto{\pgfpoint{\pgf@xa - 8pt + \pgf@shorten@left}{\pgf@ya + 1.5 * \pgf@shorten@left}}
\pgfpathlineto{\pgfpoint{\pgf@xa - 8pt + \pgf@shorten@left}{\pgf@ya}}
\pgfpathlineto{\pgfpoint{\pgf@xb + 8pt - \pgf@shorten@right}{\pgf@ya}}
\pgfpathlineto{\pgfpoint{\pgf@xb + 8pt - \pgf@shorten@right}{\pgf@ya + 1.5 * \pgf@shorten@right}}
\pgfpathclose
}
}

\makeatother

\pgfkeyssetvalue{/tikz/shorten left}{0pt}
\pgfkeyssetvalue{/tikz/shorten right}{0pt}

\tikzstyle{kpoint common}=[draw,fill=white,inner sep=1pt,minimum height=4mm]
\tikzstyle{kpoint}=[shape=cornerpoint,shorten left=5pt,kpoint common]
\tikzstyle{kpoint adjoint}=[shape=cornercopoint,shorten left=5pt,kpoint common]
\tikzstyle{kpoint conjugate}=[shape=cornerpoint,shorten right=5pt,kpoint common]
\tikzstyle{kpoint transpose}=[shape=cornercopoint,shorten right=5pt,kpoint common]
\tikzstyle{kpoint symm}=[shape=cornerpoint,shorten left=5pt,shorten right=5pt,kpoint common]

\tikzstyle{kpointdag}=[kpoint adjoint]
\tikzstyle{kpointadj}=[kpoint adjoint]
\tikzstyle{kpointconj}=[kpoint conjugate]
\tikzstyle{kpointtrans}=[kpoint transpose]

\tikzstyle{dkpoint}=[kpoint,doubled]
\tikzstyle{dkpointdag}=[kpoint adjoint,doubled]
\tikzstyle{dkcopoint}=[kpoint adjoint,doubled]
\tikzstyle{dkpointadj}=[kpoint adjoint,doubled]
\tikzstyle{dkpointconj}=[kpoint conjugate,doubled]
\tikzstyle{dkpointtrans}=[kpoint transpose,doubled]

\tikzstyle{kscalar}=[kpoint common, shape=EBox, inner xsep=-1pt, inner ysep=3pt,font=\small]
\tikzstyle{kscalarconj}=[kpoint common, shape=WBox, inner xsep=-1pt, inner ysep=3pt,font=\small]

 \tikzstyle{upground}=[circuit ee IEC,thick,ground,rotate=90,scale=2.5]
 \tikzstyle{downground}=[circuit ee IEC,thick,ground,rotate=-90,scale=2.5]
 \tikzstyle{bigground}=[regular polygon,regular polygon sides=3,draw=gray,scale=0.50,inner sep=-0.5pt,minimum width=10mm,fill=gray]

\tikzstyle{arrs}=[-latex,font=\small,auto]
\tikzstyle{arrow plain}=[arrs]
\tikzstyle{arrow dashed}=[dashed,arrs]
\tikzstyle{arrow bold}=[very thick,arrs]
\tikzstyle{arrow hide}=[draw=white!0,-]
\tikzstyle{arrow reverse}=[latex-]
\tikzstyle{cdnode}=[]

\usetikzlibrary{decorations.markings}
\usetikzlibrary{shapes.geometric}
\pgfdeclarelayer{edgelayer}
\pgfdeclarelayer{nodelayer}
\pgfsetlayers{edgelayer,nodelayer,main}

\tikzstyle{none}=[inner sep=0pt]
\definecolor{hexcolor0xff0000}{rgb}{1.000,0.000,0.000}
\definecolor{hexcolor0x000000}{rgb}{0.000,0.000,0.000}
\definecolor{hexcolor0x00ff00}{rgb}{0.000,1.000,0.000}
\definecolor{hexcolor0x000000}{rgb}{0.000,0.000,0.000}
\definecolor{hexcolor0xffff00}{rgb}{1.000,1.000,0.000}
\definecolor{hexcolor0x000000}{rgb}{0.000,0.000,0.000}

\tikzstyle{rn}=[circle,fill=hexcolor0xff0000,draw=hexcolor0x000000,line width=0.8 pt]
\tikzstyle{gn}=[circle,fill=hexcolor0x00ff00,draw=hexcolor0x000000,line width=0.8 pt]
\tikzstyle{yn}=[circle,fill=hexcolor0xffff00,draw=hexcolor0x000000,line width=0.8 pt]

\tikzstyle{simple}=[-,draw=hexcolor0x000000,line width=2.000]
\tikzstyle{arrow}=[-,draw=hexcolor0x000000,postaction={decorate},decoration={markings,mark=at position .5 with {\arrow{>}}},line width=2.000]
\tikzstyle{tick}=[-,draw=hexcolor0x000000,postaction={decorate},decoration={markings,mark=at position .5 with {\draw (0,-0.1) -- (0,0.1);}},line width=2.000]

\newtheorem{thm}{Theorem}[]
\newtheorem{define}{Definition}

\newtheorem{rem}{Remark}
\newtheorem{lem}[thm]{Lemma}
\newtheorem{prop}[thm]{Proposition}

\newtheorem{ex}{Example}

\newcommand{\diageq}{\rotatebox{-45}{$\,=$}}
\newcommand{\dadj}{{\dagger}}
\newcommand{\dax}{{\sharp}}
\newcommand{\dpt}{{\dagger_{pt}}}

\newcommand{\daggerH}{\dadj}
\newcommand{\tdadj}{{\dagger}}

\newcommand{\emptydiag}{\,\tikz{\node[style=empty diagram] (x) {};}\,}

\usepackage{color}
\def\bR{\begin{color}{red}}
\def\bB{\begin{color}{blue}}
\def\bM{\begin{color}{magenta}}
\def\bC{\begin{color}{cyan}}
\def\bW{\begin{color}{white}}
\def\bBl{\begin{color}{black}}
\def\bG{\begin{color}{green}}
\def\bY{\begin{color}{yellow}}
\def\e{\end{color}}

\newcommand{\bit}{\begin{itemize}}
\newcommand{\eit}{\end{itemize}\par\noindent}
\newcommand{\ben}{\begin{enumerate}}
\newcommand{\een}{\end{enumerate}\par\noindent}

\title{Process-theoretic characterisation of the Hermitian adjoint}
\author{John H.~Selby \& Bob Coecke}

\begin{document}

\maketitle

\begin{abstract}
We show that the physical principle ``the adjoint associates to each state a `test' for that state'' fully characterises the Hermitian adjoint for pure quantum theory, therefore providing the adjoint with operational meaning beyond its standard mathematical definition. Also, we show that for general process theories, which all admit a diagrammatic representation, this physical principle induces a reflection operation.
\end{abstract}

\section{Introduction}

The process theoretic approach \cite{abramsky2004categorical, coecke2015categorical, coecke2014picturing} provides a novel perspective on the structure of quantum theory by focusing on \emph{compositionality}. Moreover, it leads to a diagrammatic representation of quantum processes, simplifying calculations and providing an intuitive way to reason about quantum theory \cite{coecke2005kindergarten, ContPhys}. Furthermore, process-theoretic ideas have been: adopted within other frameworks and corresponding reconstruction theorems \cite{HardyPicturalism, chiribella2010probabilistic, hardy2011reformulating, chiribella2011informational}; provided a natural setting for theories other than quantum theory \cite{CES} which has led to, for example, the study of connections between physical principles and computation \cite{lee2016generalised,lee2016higher}; helped solve open problems in quantum computing e.g.~\cite{DP2, BoixoHeunen, Clare}; and provided a general framework for resource theories \cite{coecke2014mathematical}.

However, one particular diagrammatic ingredient that  plays a central role in quantum theory lacks  a clear   physical interpretation, namely, the diagrammatic counterpart to the  Hermitian adjoint. Earlier work has introduced \emph{dagger} process theories \cite{AC2, selinger2007dagger} as a generalisation of the adjoint to other theories. However, this is not entirely satisfactory, raising three questions which we answer in this paper.
\begin{itemize}
\item Firstly, why should a theory have have a dagger at all? Section 4 introduces the concept of a \emph{test structure}, associating to each state a test for that state. Section 5 shows that this leads to a dagger.
\item Secondly, why is it always assumed that the dagger should be involutive? Section 5 demonstrates how this is a consequence of the definition of the test structure.
\item Finally, why is the particular dagger of interest in quantum theory the Hermitian adjoint, rather than some other dagger such as the transpose? Section 6 shows that for quantum theory the test structure forces the dagger to be the Hermitian adjoint.
\end{itemize}
To finish in section 7 we consider process theories with a notion of probabilistic mixing, and show that such theories typically fail to have a test structure.  The reason for this is that `testability' is tightly intertwined with that of  `purity'.   This provides a new perspective on what it means for states to be pure -- they are pure if they are `testable'.

\paragraph{Related Work}
The physical principle used in this work is closely related to the `logical sharpness' axiom used by Hardy in the GPT setting~\cite{hardy2011reformulating}.  A similar approach to ours was recently suggested by Chiribella \cite{Chiribella} who relates the adjoint to time-reversal symmetry. By taking this as an axiom he has shown that the adjoint must be a process theoretic dagger. This however does not uniquely pick out the Hermitian adjoint as a dagger; for example, it also allows for the transpose.

Vicary \cite{vicary2011categorical} also characterised the adjoint in process-theoretic terms.  However, he required extremely strong assumptions, namely that all dagger-limits exist, which isn't ever the case if one eliminates redundant global phases from quantum theory.

\section{Process theories\label{PTs}}

A process theory  consists of:
\ben
\item a collection of `systems',  these  could represent a physical system such as a photon, but also a mathematical object such as a vector space, or some computational notion such as a data type;
\item  a collection of  `processes', each of which has an input consisting of a set of systems and another set of systems as an output,    these  could represent a  physical process  such as  a parametric down converter (this would have a single photon as its input, and a pair of photons as  its output), but again, these processes can also be mathematical processes such as a linear transformation, or computational processes such as a sorting algorithm;
\item  a composition operation for systems and processes.
\een
 This composition operation can be best understood in terms of a diagrammatic notation  for systems and processes.  In this notation, systems are represented by labelled wires, for example:
\[
\text{single system}:\quad\begin{tikzpicture}
	\begin{pgfonlayer}{nodelayer}
		\node [style=none] (0) at (0, -1) {};
		\node [style=none] (1) at (0, 1) {};
		\node [style=none] (2) at (0.5, -0.5) {$A$};
	\end{pgfonlayer}
	\begin{pgfonlayer}{edgelayer}
		\draw (0.center) to (1.center);
	\end{pgfonlayer}
\end{tikzpicture} \ ,\qquad\quad\text{composite system}:\quad\begin{tikzpicture}
	\begin{pgfonlayer}{nodelayer}
		\node [style=none] (0) at (0, -1) {};
		\node [style=none] (1) at (0, 1) {};
		\node [style=none] (2) at (0.5, -0.5) {$A_1$};
		\node [style=none] (3) at (1.5, -1) {};
		\node [style=none] (4) at (1.5, 1) {};
		\node [style=none] (5) at (3.25, -0) {...};
		\node [style=none] (6) at (4.5, -1) {};
		\node [style=none] (7) at (4.5, 1) {};
		\node [style=none] (8) at (2, -0.5) {$A_2$};
		\node [style=none] (9) at (5, -0.5) {$A_n$};
	\end{pgfonlayer}
	\begin{pgfonlayer}{edgelayer}
		\draw (0.center) to (1.center);
		\draw (3.center) to (4.center);
		\draw (6.center) to (7.center);
	\end{pgfonlayer}
\end{tikzpicture}\ .
\]
Processes are represented as boxes which have a set of input wires at the bottom and a set of output wires at the top. This gives a sense of a `direction of time' as flowing up the page. Dropping system labels for convenience we can denote a process $f$ as, for example:
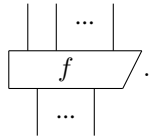
\[
\begin{tikzpicture}
	\begin{pgfonlayer}{nodelayer}
		\node [style=none] (0) at (-1.5, -0.5) {};
		\node [style=none] (1) at (1.5, -0.5) {};
		\node [style=none] (2) at (2, 0.5) {};
		\node [style=none] (3) at (-1.5, 0.5) {};
		\node [style=none] (4) at (-1, 0.5) {};
		\node [style=none] (5) at (-0.25, 0.5) {};
		\node [style=none] (6) at (1.25, 0.5) {};
		\node [style=none] (7) at (-0.75, -0.5) {};
		\node [style=none] (8) at (0.75, -0.5) {};
		\node [style=none] (9) at (-1, 1.75) {};
		\node [style=none] (10) at (-0.25, 1.75) {};
		\node [style=none] (11) at (1.25, 1.75) {};
		\node [style=none] (12) at (0.5, 1.25) {...};
		\node [style=none] (13) at (0, -1.25) {...};
		\node [style=none] (14) at (-0.75, -1.75) {};
		\node [style=none] (15) at (0.75, -1.75) {};
		\node [style=none] (16) at (0, 0) {$f$};
	\end{pgfonlayer}
	\begin{pgfonlayer}{edgelayer}
		\draw (3.center) to (2.center);
		\draw (2.center) to (1.center);
		\draw (1.center) to (0.center);
		\draw (0.center) to (3.center);
		\draw (4.center) to (9.center);
		\draw (5.center) to (10.center);
		\draw (6.center) to (11.center);
		\draw (14.center) to (7.center);
		\draw (15.center) to (8.center);
	\end{pgfonlayer}
\end{tikzpicture}.
\]
The shape of this box is irrelevant at this point but the asymmetry will be useful  later on.

There are three special types of processes: those with no input, called \em state preparations \em (or \em states \em for short); those with no outputs, called \em measurement outcomes \em (or \em effects \em for short); and those with neither inputs nor outputs, called \em scalars\em,  which  are often taken to be probabilities.
 Each of  these  has its own special notation:
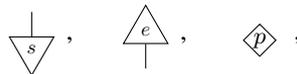
\[
\begin{tikzpicture}
	\begin{pgfonlayer}{nodelayer}
		\node [style=point] (0) at (0, -0.25) {$s$};
		\node [style=none] (1) at (0, 0.7500001) {};
		\node [style=none] (2) at (1, -0) {,};
		\node [style=none] (3) at (3, -0.75) {};
		\node [style=copoint] (4) at (3, 0.25) {$e$};
		\node [style=none] (5) at (4, -0) {,};
		\node [style=scalar] (6) at (6, -0) {$p$};
		\node [style=none] (7) at (7, -0) {,};
	\end{pgfonlayer}
	\begin{pgfonlayer}{edgelayer}
		\draw (0) to (1.center);
		\draw (3.center) to (4);
	\end{pgfonlayer}
\end{tikzpicture}
\]
 inspired by Dirac-notation \cite{coecke2005kindergarten}.
The box around a scalar is often omitted for clarity.

 We can now easily define the composition operation of a process theory:
given some collection of processes then these can be wired together to form \emph{diagrams}, for example:
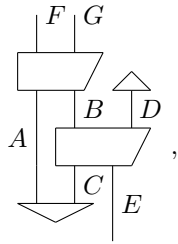
\[
\begin{tikzpicture}
	\begin{pgfonlayer}{nodelayer}
		\node [style=none] (0) at (0, -1.5) {};
		\node [style=none] (1) at (2, -1.5) {};
		\node [style=none] (2) at (0.9999999, -2) {};
		\node [style=none] (3) at (0.4999999, -1.5) {};
		\node [style=none] (4) at (0.4999999, -0.4999999) {};
		\node [style=none] (5) at (1.5, -1.5) {};
		\node [style=none] (6) at (1.5, -0.4999999) {};
		\node [style=none] (7) at (0.9999999, -0.4999999) {};
		\node [style=none] (8) at (3, -0.4999999) {};
		\node [style=none] (9) at (3.5, 0.4999999) {};
		\node [style=none] (10) at (0.9999999, 0.4999999) {};
		\node [style=none] (11) at (1.5, 0.4999999) {};
		\node [style=none] (12) at (3, 0.4999999) {};
		\node [style=none] (13) at (2.5, -0.4999999) {};
		\node [style=none] (14) at (2.5, -2.5) {};
		\node [style=none] (15) at (0.4999999, 1.5) {};
		\node [style=none] (16) at (0.4999999, 1.5) {};
		\node [style=none] (17) at (1.5, 1.5) {};
		\node [style=none] (18) at (2.5, 1.5) {};
		\node [style=none] (19) at (3, 1.5) {};
		\node [style=none] (20) at (3.5, 1.5) {};
		\node [style=none] (21) at (3, 2) {};
		\node [style=none] (22) at (1.75, 1.5) {};
		\node [style=none] (23) at (2.25, 2.5) {};
		\node [style=none] (24) at (0, 2.5) {};
		\node [style=none] (25) at (0, 1.5) {};
		\node [style=none] (26) at (0.4999999, 3.5) {};
		\node [style=none] (27) at (1.5, 3.5) {};
		\node [style=none] (28) at (0.4999999, 2.5) {};
		\node [style=none] (29) at (1.5, 2.5) {};
		\node [style=none] (30) at (0, 0.25) {$A$};
		\node [style=none] (31) at (2, 0.9999999) {$B$};
		\node [style=none] (32) at (2, -0.9999999) {$C$};
		\node [style=none] (33) at (3.5, 0.9999999) {$D$};
		\node [style=none] (34) at (3, -1.5) {$E$};
		\node [style=none] (35) at (0.9999999, 3.5) {$F$};
		\node [style=none] (36) at (2, 3.5) {$G$};
		\node [style=none] (37) at (4, -0) {};
	\end{pgfonlayer}
	\begin{pgfonlayer}{edgelayer}
		\draw (0.center) to (2.center);
		\draw (2.center) to (1.center);
		\draw (1.center) to (0.center);
		\draw (3.center) to (4.center);
		\draw (5.center) to (6.center);
		\draw (7.center) to (8.center);
		\draw (8.center) to (9.center);
		\draw (9.center) to (10.center);
		\draw (10.center) to (7.center);
		\draw (4.center) to (16.center);
		\draw (11.center) to (17.center);
		\draw (12.center) to (19.center);
		\draw (19.center) to (18.center);
		\draw (18.center) to (21.center);
		\draw (21.center) to (20.center);
		\draw (20.center) to (19.center);
		\draw (14.center) to (13.center);
		\draw (25.center) to (22.center);
		\draw (22.center) to (23.center);
		\draw (23.center) to (24.center);
		\draw (24.center) to (25.center);
		\draw (28.center) to (26.center);
		\draw (29.center) to (27.center);
	\end{pgfonlayer}
\end{tikzpicture},
\]
 which are also  processes in the theory.   This composition is  subject to the condition that when forming diagrams, any two systems wired together have to be (of) the same (type).  Essentially,  processes in a process theory are closed under forming diagrams.

 In order to describe `wiring processes together' explicitly, in particular when the processes are represented by means of standard mathematical models, it is often convenient to consider the following two primitive forms of  composition,  \em sequential composition \em and \em parallel composition\em:
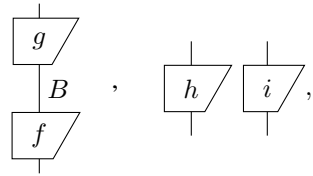
\[
\begin{tikzpicture}
	\begin{pgfonlayer}{nodelayer}
		\node [style=none] (0) at (0.5, -2) {};
		\node [style=map] (1) at (0, -1.25) {$f$};
		\node [style=map] (2) at (0, 1.25) {$g$};
		\node [style=none] (3) at (0, 2.25) {};
		\node [style=none] (4) at (0, -2.25) {};
		\node [style=none] (5) at (0.5, -0) {$B$};
		\node [style=none] (6) at (0.5, 2) {};
		\node [style=none] (7) at (2, -0) {,};
		\node [style=map] (8) at (4, -0) {$h$};
		\node [style=map] (9) at (6, -0) {$i$};
		\node [style=none] (10) at (4, 1.25) {};
		\node [style=none] (11) at (6, 1.25) {};
		\node [style=none] (12) at (4, -1.25) {}; 
		\node [style=none] (13) at (6, -1.25) {};
	\end{pgfonlayer}
	\begin{pgfonlayer}{edgelayer}
		\draw (4.center) to (1);
		\draw (1) to (2);
		\draw (2) to (3.center);
		\draw (12.center) to (8);
		\draw (8) to (10.center);
		\draw (13.center) to (9);
		\draw (9) to (11.center);
	\end{pgfonlayer}
\end{tikzpicture},
\]
 which we symbolically  denote by $g\circ f$ and $h\otimes i$ respectively.

 There are also some special scalars for a  process theory that are fully characterised by their behaviour under composition. One special scalar is `certain', which is either written as 1 or by the empty diagram, which, of course, when composed with any other process leaves that process invariant:
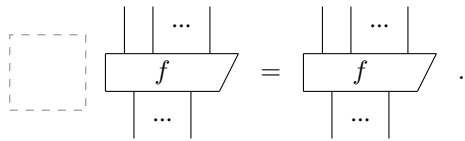
\[
\emptydiag\ \ \begin{tikzpicture}
	\begin{pgfonlayer}{nodelayer}
		\node [style=none] (0) at (-1.5, -0.5) {};
		\node [style=none] (1) at (1.5, -0.5) {};
		\node [style=none] (2) at (2, 0.5) {};
		\node [style=none] (3) at (-1.5, 0.5) {};
		\node [style=none] (4) at (-1, 0.5) {};
		\node [style=none] (5) at (-0.25, 0.5) {};
		\node [style=none] (6) at (1.25, 0.5) {};
		\node [style=none] (7) at (-0.75, -0.5) {};
		\node [style=none] (8) at (0.75, -0.5) {};
		\node [style=none] (9) at (-1, 1.75) {};
		\node [style=none] (10) at (-0.25, 1.75) {};
		\node [style=none] (11) at (1.25, 1.75) {};
		\node [style=none] (12) at (0.5, 1.25) {...};
		\node [style=none] (13) at (0, -1.25) {...};
		\node [style=none] (14) at (-0.75, -1.75) {};
		\node [style=none] (15) at (0.75, -1.75) {};
		\node [style=none] (16) at (0, 0) {$f$};
	\end{pgfonlayer}
	\begin{pgfonlayer}{edgelayer}
		\draw (3.center) to (2.center);
		\draw (2.center) to (1.center);
		\draw (1.center) to (0.center);
		\draw (0.center) to (3.center);
		\draw (4.center) to (9.center);
		\draw (5.center) to (10.center);
		\draw (6.center) to (11.center);
		\draw (14.center) to (7.center);
		\draw (15.center) to (8.center);
	\end{pgfonlayer}
\end{tikzpicture} \ \ = \ \ \begin{tikzpicture}
	\begin{pgfonlayer}{nodelayer}
		\node [style=none] (0) at (-1.5, -0.5) {};
		\node [style=none] (1) at (1.5, -0.5) {};
		\node [style=none] (2) at (2, 0.5) {};
		\node [style=none] (3) at (-1.5, 0.5) {};
		\node [style=none] (4) at (-1, 0.5) {};
		\node [style=none] (5) at (-0.25, 0.5) {};
		\node [style=none] (6) at (1.25, 0.5) {};
		\node [style=none] (7) at (-0.75, -0.5) {};
		\node [style=none] (8) at (0.75, -0.5) {};
		\node [style=none] (9) at (-1, 1.75) {};
		\node [style=none] (10) at (-0.25, 1.75) {};
		\node [style=none] (11) at (1.25, 1.75) {};
		\node [style=none] (12) at (0.5, 1.25) {...};
		\node [style=none] (13) at (0, -1.25) {...};
		\node [style=none] (14) at (-0.75, -1.75) {};
		\node [style=none] (15) at (0.75, -1.75) {};
		\node [style=none] (16) at (0, 0) {$f$};
	\end{pgfonlayer}
	\begin{pgfonlayer}{edgelayer}
		\draw (3.center) to (2.center);
		\draw (2.center) to (1.center);
		\draw (1.center) to (0.center);
		\draw (0.center) to (3.center);
		\draw (4.center) to (9.center);
		\draw (5.center) to (10.center);
		\draw (6.center) to (11.center);
		\draw (14.center) to (7.center);
		\draw (15.center) to (8.center);
	\end{pgfonlayer}
\end{tikzpicture}\ \ \ .
\]
Another one is `impossible',  which is written as 0, and `eats' all other diagrams, in the sense that for each set of input and output wires there is a 0-process, again simply denoted by 0, and when composing any process with the 0-scalar we obtain the corresponding 0-process:
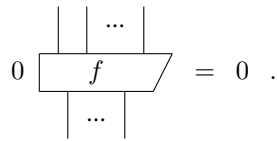
\[
0\ \ \begin{tikzpicture}
	\begin{pgfonlayer}{nodelayer}
		\node [style=none] (0) at (-1.5, -0.5) {};
		\node [style=none] (1) at (1.5, -0.5) {};
		\node [style=none] (2) at (2, 0.5) {};
		\node [style=none] (3) at (-1.5, 0.5) {};
		\node [style=none] (4) at (-1, 0.5) {};
		\node [style=none] (5) at (-0.25, 0.5) {};
		\node [style=none] (6) at (1.25, 0.5) {};
		\node [style=none] (7) at (-0.75, -0.5) {};
		\node [style=none] (8) at (0.75, -0.5) {};
		\node [style=none] (9) at (-1, 1.75) {};
		\node [style=none] (10) at (-0.25, 1.75) {};
		\node [style=none] (11) at (1.25, 1.75) {};
		\node [style=none] (12) at (0.5, 1.25) {...};
		\node [style=none] (13) at (0, -1.25) {...};
		\node [style=none] (14) at (-0.75, -1.75) {};
		\node [style=none] (15) at (0.75, -1.75) {};
		\node [style=none] (16) at (0, 0) {$f$};
	\end{pgfonlayer}
	\begin{pgfonlayer}{edgelayer}
		\draw (3.center) to (2.center);
		\draw (2.center) to (1.center);
		\draw (1.center) to (0.center);
		\draw (0.center) to (3.center);
		\draw (4.center) to (9.center);
		\draw (5.center) to (10.center);
		\draw (6.center) to (11.center);
		\draw (14.center) to (7.center);
		\draw (15.center) to (8.center);
	\end{pgfonlayer}
\end{tikzpicture} \ \ = \ \ 0 \ \ \ .
\]

More details on this process-theoretic framework can be found in \cite{coecke2015categorical, coecke2014picturing}.

\begin{ex}\label{CLM}\em
In the process theory $\mathds{C}LM$ the systems are finite dimensional complex vector spaces and the processes are linear maps between these vector spaces.
Sequential composition is composition of linear maps and parallel composition is the tensor product. State preparations can be identified with complex vectors,  effects with covectors,  and  scalars  with  the complex numbers.  This `mathematical' process theory will now be used to construct  quantum theory as a process theory.
\end{ex}

\begin{ex}
 We can construct the process theory of \underline{pure post-selected} quantum processes directly from the process theory $\mathds{C}LM$.  One way to do so is to consider equivalence classes of linear maps, by ignoring global phases.  A more elegant manner of doing the same is by means of \em doubling\em, that is, every process is replaced by its double:
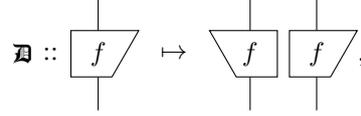
\[\begin{tikzpicture}
	\begin{pgfonlayer}{nodelayer}
		\node [style=map] (0) at (0, 0) {$f$};
		\node [style=none] (1) at (0, -1.5) {};
		\node [style=none] (2) at (0, 1.5) {};
		\node [style=none] (3) at (-2, 0) {$\textgoth{D}$};
		\node [style=none] (4) at (2, 0) {$\mapsto$};
		\node [style=none] (5) at (4, -1.5) {};
		\node [style=mapconj] (6) at (4, 0) {$f$};
		\node [style=none] (7) at (4, 1.5) {};
		\node [style=none] (8) at (5.75, -1.5) {};
		\node [style=map] (9) at (5.75, -0) {$f$};
		\node [style=none] (10) at (5.75, 1.5) {};
		\node [style=none] (11) at (-1.25, -0) {\emph{::}};
	\end{pgfonlayer}
	\begin{pgfonlayer}{edgelayer}
		\draw (1.center) to (0);
		\draw (0) to (2.center);
		\draw (5.center) to (6);
		\draw (6) to (7.center);
		\draw (8.center) to (9);
		\draw (9) to (10.center);
	\end{pgfonlayer}
\end{tikzpicture},\]
where,
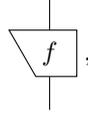
\[\begin{tikzpicture}
	\begin{pgfonlayer}{nodelayer}
		\node [style=none] (0) at (0, -1.5) {};
		\node [style=mapconj] (1) at (0, 0) {$f$};
		\node [style=none] (2) at (0, 1.5) {};
	\end{pgfonlayer}
	\begin{pgfonlayer}{edgelayer}
		\draw (0.center) to (1);
		\draw (1) to (2.center);
	\end{pgfonlayer}
\end{tikzpicture}\ ,\]
 represents the conjugate of $f$, hence obtaining a new process theory $\textgoth{D}[\mathds{C}LM]$ \cite{DeLL}.  Composing a state and an effect  now  gives a positive real number corresponding to the probability of, for example, measuring the effect $\phi$ given state $\psi$ according to the Born-rule:
\[
\begin{tikzpicture}
	\begin{pgfonlayer}{nodelayer}
		\node [style=kpointtrans] (0) at (0, 0.75) {$\phi$};
		\node [style=kpointconj] (1) at (0, -0.75) {$\psi$};
		\node [style=kpoint] (2) at (1.25, -0.75) {$\psi$};
		\node [style=kpointadj] (3) at (1.25, 0.75) {$\phi$};
		\node [style=none] (5) at (4.75, -0) {$=\ c^* c \in \mathds{R}^+$};
	\end{pgfonlayer}
	\begin{pgfonlayer}{edgelayer}
		\draw (1) to (0);
		\draw (2) to (3);
	\end{pgfonlayer}
\end{tikzpicture} \  .
\]
 Clearly, processes differing by only a global phase  become identical,
\[
\textgoth{D}(e^{i\theta}f)=e^{i\theta}e^{-i\theta}f^*\otimes f = \textgoth{D}(f).
\]
Perhaps the most obvious description of  $\textgoth{D}[\mathds{C}LM]$ is  as a generalisation of `Dirac' notation,  when representing states by ket-bras  \cite{coecke2005kindergarten}.
\end{ex}

\begin{rem}  \em
 Symbolically, in category-theoretic terms, process theories can be defined as strict symmetric monoidal categories  (SMCs). The systems of the process theory are the objects in the category, the processes are morphisms, and the states for some object are the morphisms from the tensor  unit  into that object \cite{CatsII, coecke2014picturing}.
\end{rem}

\section{Types of  process theories}

 For physical process theories we will always assume that the scalars are the positive real numbers, where the interval $[0,1]$ corresponds to `probabilistic weights'.

Inspired by the physical notion of process tomography, and given the interpretations of states and effects, we can define the following special kind of process theories:

\begin{define} \textbf{Tomographic process theories}\label{PhysPTs}
 are process theories
where all processes can be characterised in terms of scalars:
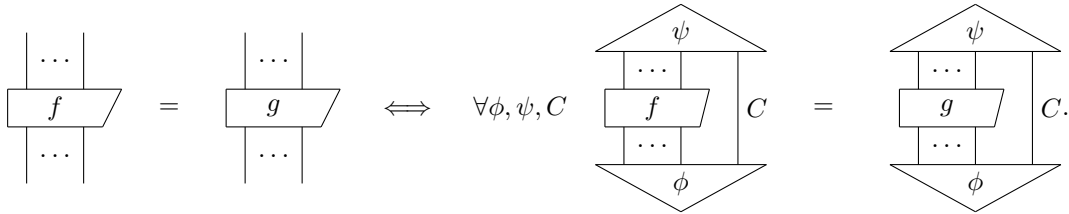
\[
\begin{tikzpicture}
	\begin{pgfonlayer}{nodelayer}
		\node [style=none] (0) at (-0.7500001, -2) {};
		\node [style=none] (1) at (-0.7500001, 2) {};
		\node [style=none] (2) at (1.5, -0) {$=$};
		\node [style=none] (3) at (0.25, 0.5) {};
		\node [style=none] (4) at (-0.25, -0.5) {};
		\node [style=none] (5) at (-2.75, -0.5) {};
		\node [style=none] (6) at (-2.75, 0.5) {};
		\node [style=none] (7) at (-0.75, 0.5) {};
		\node [style=none] (8) at (-2.25, 0.5) {};
		\node [style=none] (9) at (-2.25, -0.5) {};
		\node [style=none] (10) at (-2.25, -2) {};
		\node [style=none] (11) at (-2.25, 2) {};
		\node [style=none] (12) at (-0.75, -0.5) {};
		\node [style=none] (13) at (-1.5, -0) {$f$};
		\node [style=none] (14) at (3.5, -0.5) {};
		\node [style=none] (15) at (5, 2) {};
		\node [style=none] (16) at (5, -0.5) {};
		\node [style=none] (17) at (3, -0.5) {};
		\node [style=none] (18) at (3.5, -2) {};
		\node [style=none] (19) at (5.5, -0.5) {};
		\node [style=none] (20) at (6, 0.5) {};
		\node [style=none] (21) at (4.25, -0) {$g$};
		\node [style=none] (22) at (5, -2) {};
		\node [style=none] (23) at (3.5, 2) {};
		\node [style=none] (24) at (5, 0.5) {};
		\node [style=none] (25) at (3, 0.5) {};
		\node [style=none] (26) at (3.5, 0.5) {};
		\node [style=none] (27) at (-1.5, 1.25) {$\ldots$};
		\node [style=none] (28) at (-1.5, -1.25) {$\ldots$};
		\node [style=none] (29) at (4.25, 1.25) {$\ldots$};
		\node [style=none] (30) at (4.25, -1.25) {$\ldots$};
	\end{pgfonlayer}
	\begin{pgfonlayer}{edgelayer}
		\draw (6.center) to (3.center);
		\draw (3.center) to (4.center);
		\draw (4.center) to (5.center);
		\draw (5.center) to (6.center);
		\draw (8.center) to (11.center);
		\draw (1.center) to (7.center);
		\draw (9.center) to (10.center);
		\draw (0.center) to (12.center);
		\draw (25.center) to (20.center);
		\draw (20.center) to (19.center);
		\draw (19.center) to (17.center);
		\draw (17.center) to (25.center);
		\draw (26.center) to (23.center);
		\draw (15.center) to (24.center);
		\draw (14.center) to (18.center);
		\draw (22.center) to (16.center);
	\end{pgfonlayer}
\end{tikzpicture} \quad \iff   \quad \forall \phi,\psi, C \quad  \begin{tikzpicture}
	\begin{pgfonlayer}{nodelayer}
		\node [style=none] (0) at (-0.5, -0) {$=$};
		\node [style=none] (1) at (-3.5, 0.5) {};
		\node [style=none] (2) at (-3.75, -0.5) {};
		\node [style=none] (3) at (-6.25, -0.5) {};
		\node [style=none] (4) at (-6.25, 0.5) {};
		\node [style=none] (5) at (-4.25, 0.5) {};
		\node [style=none] (6) at (-5.75, 0.5) {};
		\node [style=none] (7) at (-5.75, -0.5) {};
		\node [style=none] (8) at (-4.25, -0.5) {};
		\node [style=none] (9) at (-5, -0) {$f$};
		\node [style=none] (10) at (-5.75, 1.5) {};
		\node [style=none] (11) at (-4.25, 1.5) {};
		\node [style=none] (12) at (-4.25, -1.5) {};
		\node [style=none] (13) at (-5.75, -1.5) {};
		\node [style=none] (14) at (-2.75, -1.5) {};
		\node [style=none] (15) at (-2.75, 1.5) {};
		\node [style=none] (16) at (-2, 1.5) {};
		\node [style=none] (17) at (-6.5, 1.5) {};
		\node [style=none] (18) at (-6.5, -1.5) {};
		\node [style=none] (19) at (-2, -1.5) {};
		\node [style=none] (20) at (-4.25, -2.75) {};
		\node [style=none] (21) at (-4.25, 2.75) {};
		\node [style=none] (22) at (-4.25, 2) {$\psi$};
		\node [style=none] (23) at (-4.25, -2) {$\phi$};
		\node [style=none] (24) at (-2.25, -0) {$C$};
		\node [style=none] (25) at (2, -0.5) {};
		\node [style=none] (26) at (5.75, 1.5) {};
		\node [style=none] (27) at (3.5, -2) {$\phi$};
		\node [style=none] (28) at (2, -1.5) {};
		\node [style=none] (29) at (3.5, -2.75) {};
		\node [style=none] (30) at (5.5, -0) {$C$};
		\node [style=none] (31) at (2.75, -0) {$g$};
		\node [style=none] (32) at (1.25, 1.5) {};
		\node [style=none] (33) at (3.5, 0.5) {};
		\node [style=none] (34) at (3.5, 2.75) {};
		\node [style=none] (35) at (5, -1.5) {};
		\node [style=none] (36) at (4.25, 0.5) {};
		\node [style=none] (37) at (1.5, 0.5) {};
		\node [style=none] (38) at (2, 1.5) {};
		\node [style=none] (39) at (3.5, -0.5) {};
		\node [style=none] (40) at (1.25, -1.5) {};
		\node [style=none] (41) at (4, -0.5) {};
		\node [style=none] (42) at (3.5, 2) {$\psi$};
		\node [style=none] (43) at (2, 0.5) {};
		\node [style=none] (44) at (5, 1.5) {};
		\node [style=none] (45) at (3.5, -1.5) {};
		\node [style=none] (46) at (5.75, -1.5) {};
		\node [style=none] (47) at (1.5, -0.5) {};
		\node [style=none] (48) at (3.5, 1.5) {};
		\node [style=none] (49) at (-5, 0.9999998) {$\ldots$};
		\node [style=none] (50) at (-5, -0.9999998) {$\ldots$};
		\node [style=none] (51) at (2.75, 0.9999998) {$\ldots$};
		\node [style=none] (52) at (2.75, -0.9999998) {$\ldots$};
	\end{pgfonlayer}
	\begin{pgfonlayer}{edgelayer}
		\draw (4.center) to (1.center);
		\draw (1.center) to (2.center);
		\draw (2.center) to (3.center);
		\draw (3.center) to (4.center);
		\draw (17.center) to (16.center);
		\draw (16.center) to (21.center);
		\draw (21.center) to (17.center);
		\draw (18.center) to (19.center);
		\draw (19.center) to (20.center);
		\draw (20.center) to (18.center);
		\draw (13.center) to (7.center);
		\draw (6.center) to (10.center);
		\draw (11.center) to (5.center);
		\draw (8.center) to (12.center);
		\draw (14.center) to (15.center);
		\draw (37.center) to (36.center);
		\draw (36.center) to (41.center);
		\draw (41.center) to (47.center);
		\draw (47.center) to (37.center);
		\draw (32.center) to (26.center);
		\draw (26.center) to (34.center);
		\draw (34.center) to (32.center);
		\draw (40.center) to (46.center);
		\draw (46.center) to (29.center);
		\draw (29.center) to (40.center);
		\draw (28.center) to (25.center);
		\draw (43.center) to (38.center);
		\draw (48.center) to (33.center);
		\draw (39.center) to (45.center);
		\draw (35.center) to (44.center);
	\end{pgfonlayer}
\end{tikzpicture}.
\]
\textbf{Locally tomographic process theories}
 are tomographic process theories which satisfy the stronger condition of `local process tomography'~\cite{Araki1980,Bergia1980,wootters1990local,chiribella2011informational}:
\[
\begin{tikzpicture}
	\begin{pgfonlayer}{nodelayer}
		\node [style=none] (0) at (-0.7500001, -2) {};
		\node [style=none] (1) at (-0.7500001, 2) {};
		\node [style=none] (2) at (1.5, -0) {$=$};
		\node [style=none] (3) at (0.25, 0.5) {};
		\node [style=none] (4) at (-0.25, -0.5) {};
		\node [style=none] (5) at (-2.75, -0.5) {};
		\node [style=none] (6) at (-2.75, 0.5) {};
		\node [style=none] (7) at (-0.75, 0.5) {};
		\node [style=none] (8) at (-2.25, 0.5) {};
		\node [style=none] (9) at (-2.25, -0.5) {};
		\node [style=none] (10) at (-2.25, -2) {};
		\node [style=none] (11) at (-2.25, 2) {};
		\node [style=none] (12) at (-0.75, -0.5) {};
		\node [style=none] (13) at (-1.5, -0) {$f$};
		\node [style=none] (14) at (3.5, -0.5) {};
		\node [style=none] (15) at (5, 2) {};
		\node [style=none] (16) at (5, -0.5) {};
		\node [style=none] (17) at (3, -0.5) {};
		\node [style=none] (18) at (3.5, -2) {};
		\node [style=none] (19) at (5.5, -0.5) {};
		\node [style=none] (20) at (6, 0.5) {};
		\node [style=none] (21) at (4.25, -0) {$g$};
		\node [style=none] (22) at (5, -2) {};
		\node [style=none] (23) at (3.5, 2) {};
		\node [style=none] (24) at (5, 0.5) {};
		\node [style=none] (25) at (3, 0.5) {};
		\node [style=none] (26) at (3.5, 0.5) {};
		\node [style=none] (27) at (-1.5, 1.25) {$\ldots$};
		\node [style=none] (28) at (-1.5, -1.25) {$\ldots$};
		\node [style=none] (29) at (4.25, 1.25) {$\ldots$};
		\node [style=none] (30) at (4.25, -1.25) {$\ldots$};
	\end{pgfonlayer}
	\begin{pgfonlayer}{edgelayer}
		\draw (6.center) to (3.center);
		\draw (3.center) to (4.center);
		\draw (4.center) to (5.center);
		\draw (5.center) to (6.center);
		\draw (8.center) to (11.center);
		\draw (1.center) to (7.center);
		\draw (9.center) to (10.center);
		\draw (0.center) to (12.center);
		\draw (25.center) to (20.center);
		\draw (20.center) to (19.center);
		\draw (19.center) to (17.center);
		\draw (17.center) to (25.center);
		\draw (26.center) to (23.center);
		\draw (15.center) to (24.center);
		\draw (14.center) to (18.center);
		\draw (22.center) to (16.center);
	\end{pgfonlayer}
\end{tikzpicture} 
 \quad \iff \quad \forall \{\phi_i\},\{\psi_j\}  \quad
\begin{tikzpicture}
	\begin{pgfonlayer}{nodelayer}
		\node [style=point] (0) at (-0.4999996, -2) {$\psi_n$};
		\node [style=copoint] (1) at (-0.4999996, 2) {$\phi_m$};
		\node [style=none] (2) at (1.75, -0) {$=$};
		\node [style=none] (3) at (0.4999996, 0.5000003) {};
		\node [style=none] (4) at (0, -0.5000003) {};
		\node [style=none] (5) at (-3, -0.5000003) {};
		\node [style=none] (6) at (-3, 0.5000003) {};
		\node [style=none] (7) at (-0.4999996, 0.5000003) {};
		\node [style=none] (8) at (-2.500001, 0.5000003) {};
		\node [style=none] (9) at (-2.500001, -0.5000003) {};
		\node [style=point] (10) at (-2.500001, -2) {$\psi_1$};
		\node [style=copoint] (11) at (-2.500001, 2) {$\phi_1$};
		\node [style=none] (12) at (-0.4999996, -0.5000003) {};
		\node [style=none] (13) at (-1.5, -0) {$f$};
		\node [style=none] (14) at (3.25, -0.5) {};
		\node [style=none] (15) at (3.25, 0.5) {};
		\node [style=none] (16) at (5.75, 0.5000003) {};
		\node [style=none] (17) at (3.75, 0.5) {};
		\node [style=point] (18) at (3.75, -2) {$\psi_1$};
		\node [style=none] (19) at (4.75, -0) {$g$};
		\node [style=none] (20) at (5.75, -0.5000003) {};
		\node [style=none] (21) at (6.25, -0.5000003) {};
		\node [style=point] (22) at (5.75, -2) {$\psi_n$};
		\node [style=none] (23) at (6.749999, 0.5000003) {};
		\node [style=copoint] (24) at (5.75, 2) {$\phi_m$};
		\node [style=copoint] (25) at (3.75, 2) {$\phi_1$};
		\node [style=none] (26) at (3.75, -0.5) {};
		\node [style=none] (27) at (-1.5, 0.9999998) {$\ldots$};
		\node [style=none] (28) at (-1.5, -0.9999998) {$\ldots$};
		\node [style=none] (29) at (4.75, 0.9999998) {$\ldots$};
		\node [style=none] (30) at (4.75, -0.9999998) {$\ldots$};
	\end{pgfonlayer}
	\begin{pgfonlayer}{edgelayer}
		\draw (6.center) to (3.center);
		\draw (3.center) to (4.center);
		\draw (4.center) to (5.center);
		\draw (5.center) to (6.center);
		\draw (8.center) to (11);
		\draw (1) to (7.center);
		\draw (9.center) to (10);
		\draw (0) to (12.center);
		\draw (15.center) to (23.center);
		\draw (23.center) to (21.center);
		\draw (21.center) to (14.center);
		\draw (14.center) to (15.center);
		\draw (17.center) to (25);
		\draw (24) to (16.center);
		\draw (26.center) to (18);
		\draw (22) to (20.center);
	\end{pgfonlayer}
\end{tikzpicture} .
\]
\end{define}

\begin{ex}
Quantum theory, in the form of $\textgoth{D}[\mathds{C}LM]$ is a locally tomographic process theory \cite{hardy2011reformulating, coecke2014picturing}.
\end{ex}

Previous work on generalising the Hermitian adjoint to the process theoretic setting led to the notion of dagger process theories.
\begin{define} \textbf{Dagger process theories} \label{DagPTs}
are process theories which come with a `dagger', $\dpt$, an operation that assigns to each processes another process as its reflection:
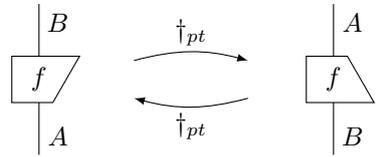
\[
\begin{tikzpicture}
	\begin{pgfonlayer}{nodelayer}
		\node [style=map] (0) at (-4, 0) {$f$};
		\node [style=mapdag] (1) at (3.75, 0) {$f$};
		\node [style=none] (2) at (-4, 2) {};
		\node [style=none] (3) at (3.75, 2) {};
		\node [style=none] (4) at (-4, -2) {};
		\node [style=none] (5) at (3.75, -2) {};
		\node [style=none] (6) at (1.5, -0.5) {};
		\node [style=none] (7) at (1.5, 0.5) {};
		\node [style=none] (8) at (0, 1.25) {$\dpt$};
		\node [style=none] (9) at (-1.5, -0.5) {};
		\node [style=none] (10) at (0, -1.25) {$\dpt$};
		\node [style=none] (11) at (-1.5, 0.5) {};
		\node [style=none] (12) at (-3.5, -1.5) {$A$};
		\node [style=none] (13) at (4.25, 1.5) {$A$};
		\node [style=none] (14) at (-3.5, 1.5) {$B$};
		\node [style=none] (15) at (4.25, -1.5) {$B$};
	\end{pgfonlayer}
	\begin{pgfonlayer}{edgelayer}
		\draw (4.center) to (0);
		\draw (0) to (2.center);
		\draw (5.center) to (1);
		\draw (1) to (3.center);
		\draw [style=arrow plain, bend left=15, looseness=1.00] (11.center) to (7.center);
		\draw [style=arrow plain, bend left=15, looseness=1.00] (6.center) to (9.center);
	\end{pgfonlayer}
\end{tikzpicture}\ ,
\]
which moreover preserves diagrams, that is, diagrams are reflected `as a whole':
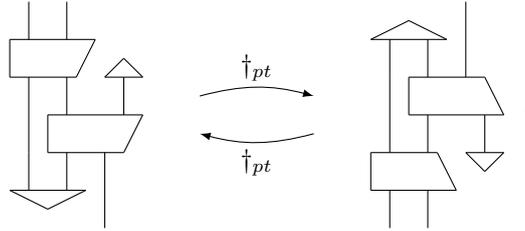
\[
\begin{tikzpicture}
	\begin{pgfonlayer}{nodelayer}
		\node [style=none] (0) at (-6.5, -2.25) {};
		\node [style=none] (1) at (-4.5, -2.25) {};
		\node [style=none] (2) at (-5.5, -2.75) {};
		\node [style=none] (3) at (-6, -2.25) {};
		\node [style=none] (4) at (-6, -1.25) {};
		\node [style=none] (5) at (-5, -2.25) {};
		\node [style=none] (6) at (-5, -1.25) {};
		\node [style=none] (7) at (-5.5, -1.25) {};
		\node [style=none] (8) at (-3.5, -1.25) {};
		\node [style=none] (9) at (-3, -0.25) {};
		\node [style=none] (10) at (-5.5, -0.25) {};
		\node [style=none] (11) at (-5, -0.25) {};
		\node [style=none] (12) at (-3.5, -0.25) {};
		\node [style=none] (13) at (-4, -1.25) {};
		\node [style=none] (14) at (-4, -3.25) {};
		\node [style=none] (15) at (-6, 0.75) {};
		\node [style=none] (16) at (-6, 0.75) {};
		\node [style=none] (17) at (-5, 0.75) {};
		\node [style=none] (18) at (-4, 0.75) {};
		\node [style=none] (19) at (-3.5, 0.75) {};
		\node [style=none] (20) at (-3, 0.75) {};
		\node [style=none] (21) at (-3.5, 1.25) {};
		\node [style=none] (22) at (-4.75, 0.75) {};
		\node [style=none] (23) at (-4.25, 1.75) {};
		\node [style=none] (24) at (-6.5, 1.75) {};
		\node [style=none] (25) at (-6.5, 0.75) {};
		\node [style=none] (26) at (-6, 2.75) {};
		\node [style=none] (27) at (-5, 2.75) {};
		\node [style=none] (28) at (-6, 1.75) {};
		\node [style=none] (29) at (-5, 1.75) {};
		\node [style=none] (30) at (3.5, -1.25) {};
		\node [style=none] (31) at (4, 2.25) {};
		\node [style=none] (32) at (3.5, 0.75) {};
		\node [style=none] (33) at (6.5, -0.25) {};
		\node [style=none] (34) at (3.5, -2.25) {};
		\node [style=none] (35) at (4.5, -2.25) {};
		\node [style=none] (36) at (5, 1.75) {};
		\node [style=none] (37) at (5.5, 2.75) {};
		\node [style=none] (38) at (6, 0.75) {};
		\node [style=none] (39) at (4.5, -0.25) {};
		\node [style=none] (40) at (4.75, -1.25) {};
		\node [style=none] (41) at (3.5, -3.25) {};
		\node [style=none] (42) at (3, 1.75) {};
		\node [style=none] (43) at (5.5, 0.75) {};
		\node [style=none] (44) at (4.5, 1.75) {};
		\node [style=none] (45) at (5.25, -2.25) {};
		\node [style=none] (46) at (6, -0.25) {};
		\node [style=none] (47) at (5.5, -1.25) {};
		\node [style=none] (48) at (4.5, 0.75) {};
		\node [style=none] (49) at (6, -1.75) {};
		\node [style=none] (50) at (3, -2.25) {};
		\node [style=none] (51) at (4, -0.25) {};
		\node [style=none] (52) at (3, -1.25) {};
		\node [style=none] (53) at (3.5, -1.25) {};
		\node [style=none] (54) at (6, -1.25) {};
		\node [style=none] (55) at (3.5, 1.75) {};
		\node [style=none] (56) at (4.5, -1.25) {};
		\node [style=none] (57) at (4, 0.75) {};
		\node [style=none] (58) at (6.5, -1.25) {};
		\node [style=none] (59) at (4.5, -3.25) {};
		\node [style=none] (60) at (-1.5, 0.25) {};
		\node [style=none] (61) at (1.5, 0.25) {};
		\node [style=none] (62) at (1.5, -0.75) {};
		\node [style=none] (63) at (-1.5, -0.75) {};
		\node [style=none] (64) at (0, 1) {$\dpt$};
		\node [style=none] (65) at (0, -1.5) {$\dpt$};
		\node [style=none] (66) at (7, -0) {};
	\end{pgfonlayer}
	\begin{pgfonlayer}{edgelayer}
		\draw (0.center) to (2.center);
		\draw (2.center) to (1.center);
		\draw (1.center) to (0.center);
		\draw (3.center) to (4.center);
		\draw (5.center) to (6.center);
		\draw (7.center) to (8.center);
		\draw (8.center) to (9.center);
		\draw (9.center) to (10.center);
		\draw (10.center) to (7.center);
		\draw (4.center) to (16.center);
		\draw (11.center) to (17.center);
		\draw (12.center) to (19.center);
		\draw (19.center) to (18.center);
		\draw (18.center) to (21.center);
		\draw (21.center) to (20.center);
		\draw (20.center) to (19.center);
		\draw (14.center) to (13.center);
		\draw (25.center) to (22.center);
		\draw (22.center) to (23.center);
		\draw (23.center) to (24.center);
		\draw (24.center) to (25.center);
		\draw (28.center) to (26.center);
		\draw (29.center) to (27.center);
		\draw (42.center) to (31.center);
		\draw (31.center) to (36.center);
		\draw (36.center) to (42.center);
		\draw (55.center) to (32.center);
		\draw (44.center) to (48.center);
		\draw (57.center) to (38.center);
		\draw (38.center) to (33.center);
		\draw (33.center) to (51.center);
		\draw (51.center) to (57.center);
		\draw (32.center) to (53.center);
		\draw (39.center) to (56.center);
		\draw (46.center) to (54.center);
		\draw (54.center) to (47.center);
		\draw (47.center) to (49.center);
		\draw (49.center) to (58.center);
		\draw (58.center) to (54.center);
		\draw (37.center) to (43.center);
		\draw (52.center) to (40.center);
		\draw (40.center) to (45.center);
		\draw (45.center) to (50.center);
		\draw (50.center) to (52.center);
		\draw (34.center) to (41.center);
		\draw (35.center) to (59.center);
		\draw [style={arrow plain}, bend left=15, looseness=1.00] (60.center) to (61.center);
		\draw [style={arrow plain}, bend left=15, looseness=1.00] (62.center) to (63.center);
	\end{pgfonlayer}
\end{tikzpicture},
\]
 and which, as the notation already suggests, is involutive.
\end{define}

\begin{rem}
In category-theoretic terms, a dagger process theory is an SMC equipped with a identity-on-objects involutive contravariant monoidal  endofunctor \cite{selinger2007dagger}.
\end{rem}

\begin{ex}
 Pure quantum theory $\textgoth{D}[\mathds{C}LM]$ involves a dagger structure, namely the Hermitian adjoint.  More specifically, for the process theory $\mathds{C}LM$  underpinning  quantum theory, for a process $f:X\to Y$ it is defined by:
\[
\langle f^\dadj \psi,\phi\rangle_X = \langle \psi,f\phi\rangle_Y \quad \forall \psi,\phi.
\]
by doubling this lifts to $\textgoth{D}[\mathds{C}LM]$.
\end{ex}

 This raises an important  question:
\begin{itemize}
\item Why does it `mean' for quantum theory to have a dagger structure?
\end{itemize}
and in particular the following two sub-questions:
\begin{itemize}
\item Why does it have to be involutive?
\item Given that there are other candidate dagger structures for quantum theory, most notably the transpose, why is that central role played specifically by the Hermitian adjoint?
\end{itemize}

\section{The new physical principle}

In what sense can something be considered a state if there is no way to check that it is indeed that state?  Therefore we state the following physical principle:
\begin{center}
\emph{\textbf{ for  each state there exists a  corresponding `test'.}}
\end{center}
 In the remainder of this paper, after giving formal substance to this principle for general process theories, we will show that the assignment of tests for \underline{pure} quantum theory is given by the Hermitian adjoint,  after  showing that for general process theories this principle gives rise to a dagger.

\begin{define} \textbf{A  test structure $\dax$} is a mapping  from states to effects on the same type subject to the following conditions:

\begin{enumerate}
\item \textsf{Composability} of tests:
\[
\left(\begin{tikzpicture}
	\begin{pgfonlayer}{nodelayer}
		\node [style=point] (0) at (0, -0) {$\psi$};
		\node [style=none] (1) at (0, 1) {};
		\node [style=point] (2) at (1.5, -0) {$\phi$};
		\node [style=none] (3) at (1.5, 0.9999999) {};
	\end{pgfonlayer}
	\begin{pgfonlayer}{edgelayer}
		\draw (0) to (1.center);
		\draw (2) to (3.center);
	\end{pgfonlayer}
\end{tikzpicture} \right)^\dax \  =\ \  \ \begin{tikzpicture}
	\begin{pgfonlayer}{nodelayer}
		\node [style=none] (0) at (0, -1) {};
		\node [style=copoint] (1) at (0, 0) {$\psi^\dax$};
		\node [style=none] (2) at (1.5, -0.9999999) {};
		\node [style=copoint] (3) at (1.5, -0) {$\phi^\dax$};
	\end{pgfonlayer}
	\begin{pgfonlayer}{edgelayer}
		\draw (0.center) to (1);
		\draw (2.center) to (3);
	\end{pgfonlayer}
\end{tikzpicture}\ \ .
\]
\item \textsf{Transformability} of tests:
\[
\forall f\ \exists f^\dax\quad \text{ s.t. }\quad\left(\begin{tikzpicture}
	\begin{pgfonlayer}{nodelayer}
		\node [style=point] (0) at (0, -1.25) {$\psi$};
		\node [style=map] (1) at (0, 0.5) {$f$};
		\node [style=none] (2) at (0, 2) {};
		\node [style=none] (3) at (0.5, -0.5) {$A$};
		\node [style=none] (4) at (0.5, 1.75) {$B$};
	\end{pgfonlayer}
	\begin{pgfonlayer}{edgelayer}
		\draw[style=none] (0) to (1);
		\draw [style=none](1) to (2.center);
	\end{pgfonlayer}
\end{tikzpicture} \right)^\dax \  =\ \ \ \begin{tikzpicture}
	\begin{pgfonlayer}{nodelayer}
		\node [style=none] (0) at (0, -2.25) {};
		\node [style=map] (1) at (0, -0.75) {$f^\dax$};
		\node [style=copoint] (2) at (0, 1) {$\psi^\dax$};
		\node [style=none] (3) at (0.5, -1.75) {$B$};
		\node [style=none] (4) at (0.5, 0.25) {$A$};
	\end{pgfonlayer}
	\begin{pgfonlayer}{edgelayer}
		\draw [style=none] (0.center) to (1);
		\draw [style=none] (1) to (2);
	\end{pgfonlayer}
\end{tikzpicture}\ \ .
\]
\item \textsf{Tests produce probabilities}:
\[
 \begin{tikzpicture}
	\begin{pgfonlayer}{nodelayer}
		\node [style=none] (0) at (-11.25, 0) {If};
		\node [style=copoint] (1) at (-8.75, 0.4999999) {$\psi^\dax$};
		\node [style=point] (2) at (-8.75, -0.4999999) {$\psi$};
		\node [style=none] (3) at (-7.25, -0) {$=$};
		\node [style=none] (4) at (-6.5, 0) {$1$};
		\node [style=none] (5) at (-5.75, -0) {$=$};
		\node [style=point] (6) at (-4.25, -0.4999999) {$\phi$};
		\node [style=copoint] (7) at (-4.25, 0.4999999) {$\phi^\dax$};
		\node [style=none] (8) at (-1.25, 0) {then,};
	\end{pgfonlayer}
	\begin{pgfonlayer}{edgelayer}
		\draw (2) to (1);
		\draw (6) to (7);
	\end{pgfonlayer}
\end{tikzpicture}\ \quad
\begin{tikzpicture}
	\begin{pgfonlayer}{nodelayer}
		\node [style=point] (0) at (-0.9999999, -0.4999999) {$\phi$};
		\node [style=copoint] (1) at (-0.9999999, 0.4999999) {$\psi^\dax$};
		\node [style=none] (2) at (0.5, 0) {$\leq$};
		\node [style=none] (3) at (1.5, 0) {$1$};
	\end{pgfonlayer}
	\begin{pgfonlayer}{edgelayer}
		\draw (0) to (1);
	\end{pgfonlayer}
\end{tikzpicture} \ .
\]
\item \textsf{Testability} of all states:
\[
\forall\
\chi \neq 0 \quad \exists  r
\quad \text{s.t.}\quad
 \begin{tikzpicture}
	\begin{pgfonlayer}{nodelayer}
		\node [style=point] (0) at (0.25, -0.75) {$\chi$};
		\node [style=copoint] (1) at (0.2499997, 0.7499999) {$\chi^\dax$};
		\node [style=scalar] (2) at (-1, -1) {$r$};
		\node [style=scalar] (3) at (-1, 1) {$r^\dax\!$};
		\node [style=none] (4) at (-2.5, 0.2500001) {};
		\node [style=none] (5) at (-0.5000002, 2.75) {};
		\node [style=none] (6) at (1.5, 0.2500001) {};
		\node [style=none] (7) at (1.5, -0.2500001) {};
		\node [style=none] (8) at (-0.5000002, -2.75) {};
		\node [style=none] (9) at (-2.5, -0.2500001) {};
	\end{pgfonlayer}
	\begin{pgfonlayer}{edgelayer}
		\draw (0) to (1);
		\draw [dashed] (4.center) to (5.center);
		\draw [dashed] (5.center) to (6.center);
		\draw [dashed] (6.center) to (4.center);
		\draw [dashed] (9.center) to (8.center);
		\draw [dashed] (8.center) to (7.center);
		\draw [dashed] (7.center) to (9.center);
	\end{pgfonlayer}
\end{tikzpicture}
\ \ =1\ .
\]
\item \textsf{Sharpness} of tests:
\[
\begin{tikzpicture}
	\begin{pgfonlayer}{nodelayer}
		\node [style=none] (0) at (-11.25, 0) {If};
		\node [style=copoint] (1) at (-8.75, 0.4999999) {$\psi^\dax$};
		\node [style=point] (2) at (-8.75, -0.4999999) {$\psi$};
		\node [style=none] (3) at (-7.25, -0) {$=$};
		\node [style=none] (4) at (-6.5, 0) {$1$};
		\node [style=none] (5) at (-5.75, -0) {$=$};
		\node [style=point] (6) at (-4.25, -0.4999999) {$\phi$};
		\node [style=copoint] (7) at (-4.25, 0.4999999) {$\phi^\dax$};
		\node [style=none] (8) at (-1.25, 0) {then,};
		\node [style=copoint] (9) at (1.5, 0.4999999) {$\psi^\dax$};
		\node [style=point] (10) at (1.5, -0.4999999) {$\phi$};
		\node [style=none] (11) at (3, 0) {$=$};
		\node [style=none] (12) at (3.75, -0) {$1$};
		\node [style=none] (13) at (5.499999, -0) {$\iff$};
		\node [style=point] (14) at (7.75, -0.25) {$\phi$};
		\node [style=none] (15) at (9.000001, -0) {$=$};
		\node [style=point] (16) at (10.25, -0.25) {$\psi$};
		\node [style=none] (17) at (7.75, 0.7499999) {};
		\node [style=none] (18) at (10.25, 0.7499999) {};
	\end{pgfonlayer}
	\begin{pgfonlayer}{edgelayer}
		\draw (2) to (1);
		\draw (6) to (7);
		\draw (10) to (9);
		\draw (14) to (17.center);
		\draw (16) to (18.center);
	\end{pgfonlayer}
\end{tikzpicture} \ \ .
\]

\end{enumerate}
\end{define}

Clearly, \textsf{sharpness} constitutes the core of these axioms, the others mainly guaranteeing that test structures are `well-behaved' with respect the interpretation of $[0,1]$ as probabilities and the structure of process theories. Of course, for pure quantum theory (up to a global phase), there is a unique normalised effect which gives unit probability for a particular normalised state. So it may seem at first like \textsf{sharpness} trivially suffices to single out the Hermitian adjoint (or  at least, it's action on states and effects), however this is not the case, as in quantum theory normalisation of a state itself is defined in terms of the Hermitian adjoint.

Note also that in \textsf{transformability} the use of the notation $\dax$ for arbitrary processes that transform effects is justified, since when we take $f$ to be a state, it can be taken to be the test structure $\dax$.

\section{Test structures provide daggers\label{AdjPT}}

\begin{lem}\label{lem:preserves-certainty} Test structures preserves certainty, i.e.~$\dax(1)=1$. \label{certPres}\end{lem}
\proof For scalars $r_1, r_2$ \textsf{composability} implies that $\dax(r_1r_2)=\dax(r_1)\dax(r_2)$. Let $r_1=1$ then $\dax(r_2)=\dax(1)\dax(r_2)$. There are two solutions for this equation, i) $\dax(r_2)=0\ \forall r_2$ which violates \textsf{testability} leaving just ii) $\dax(1)=1$. \endproof

\begin{lem}\label{involutive}\label{lem:InvNormState}
  Test structures, extended to effects via \textsf{transformability}, are involutive on normalised states i.e.~$\psi^{\dax^\dax}=\psi$.
\end{lem}
\proof
Using \textsf{transformability} and lemma \ref{lem:preserves-certainty}, for normalised states we have,
\[
\begin{tikzpicture}
	\begin{pgfonlayer}{nodelayer}
		\node [style=none] (0) at (-2, 0) {$\dax $};
		\node [style=point] (1) at (-0.25, -0.4999999) {$\psi\ $};
		\node [style=copoint] (2) at (-0.25, 0.4999999) {$\psi^{\dax}$};
		\node [style=none] (3) at (2.25, 0) {$=$};
		\node [style=copoint] (4) at (-5.75, 0.7499999) {$\psi^{\dax\ }$};
		\node [style=point] (5) at (-5.75, -0.7499999) {$\psi^{\dax^\dax}$};
		\node [style=none] (6) at (-1.25, 1.5) {};
		\node [style=none] (7) at (-1.25, -1.25) {};
		\node [style=none] (8) at (0.7499999, -1.25) {};
		\node [style=none] (9) at (0.7499999, 1.5) {};
		\node [style=none] (10) at (3.75, 0) {$\dax(1)$};
		\node [style=none] (11) at (5.25, 0) {$=$};
		\node [style=none] (12) at (6, 0) {$1$};
		\node [style=none] (13) at (-3.5, 0) {$=$};
	\end{pgfonlayer}
	\begin{pgfonlayer}{edgelayer}
		\draw (1) to (2);
		\draw (5) to (4);
		\draw [bend left=15, looseness=1.00] (7.center) to (6.center);
		\draw [bend right=15, looseness=1.00] (8.center) to (9.center);
	\end{pgfonlayer}
\end{tikzpicture}\ \ ,
\]
then, by \textsf{sharpness},  $\psi^{\dax^\dax}=\psi$. \endproof

\begin{lem}\label{lem:InvNormEffect}
 For \underline{tomographic} process theories, a test structure is involutive on tests for normalised states, i.e. $\psi^{\dax^{\dax^\dax}}=\psi^\dax$.
\end{lem}
\proof
 For $\psi^\dax$, a test for a normalised state $\psi$, we have,
 \[\begin{tikzpicture}
	\begin{pgfonlayer}{nodelayer}
		\node [style=none] (0) at (3.75, -0) {$\dax $};
		\node [style=point] (1) at (5.499999, -0.4999999) {$\psi\ $};
		\node [style=copoint] (2) at (5.499999, 0.4999999) {$\psi^{\dax}$};
		\node [style=none] (3) at (7.999999, -0) {$=$};
		\node [style=copoint] (4) at (-5.75, 0.7499999) {$\psi^{\dax^{\dax^\dax}}$};
		\node [style=point] (5) at (-5.75, -0.7499999) {$\hspace{1.5mm} \psi\hspace{1.5mm} $};
		\node [style=none] (6) at (4.5, 1.5) {};
		\node [style=none] (7) at (4.5, -1.25) {};
		\node [style=none] (8) at (6.499999, -1.25) {};
		\node [style=none] (9) at (6.499999, 1.5) {};
		\node [style=none] (10) at (9.5, -0) {$\dax(\dax(1))$};
		\node [style=none] (11) at (11, -0) {$=$};
		\node [style=none] (12) at (11.75, -0) {$1$};
		\node [style=none] (13) at (-3.5, 0) {$=$};
		\node [style=point] (14) at (-0.7499999, -0.7499999) {$\psi^{\dax^{\dax}}$};
		\node [style=copoint] (15) at (-0.7499999, 0.7499999) {$\psi^{\dax^{\dax^\dax}}$};
		\node [style=none] (16) at (1.75, -0) {$=$};
		\node [style=none] (17) at (2.75, -0) {$\dax $};
		\node [style=none] (18) at (3.5, -1.25) {};
		\node [style=none] (19) at (3.5, 1.5) {};
		\node [style=none] (20) at (6.999999, -1.25) {};
		\node [style=none] (21) at (6.999999, 1.5) {};
	\end{pgfonlayer}
	\begin{pgfonlayer}{edgelayer}
		\draw (1) to (2);
		\draw (5) to (4);
		\draw [bend left=15, looseness=1.00] (7.center) to (6.center);
		\draw [bend right=15, looseness=1.00] (8.center) to (9.center);
		\draw (14) to (15);
		\draw [bend left=15, looseness=1.00] (18.center) to (19.center);
		\draw [bend right=15, looseness=1.00] (20.center) to (21.center);
	\end{pgfonlayer}
\end{tikzpicture}\ \ ,\]
using lemma~\ref{lem:InvNormState} for the first equality and \textsf{transformability} twice for the second. Then, by \textsf{sharpness}, $\psi^{\dax^{\dax^\dax}}=\psi^\dax$.
\endproof

\begin{lem} \label{lem:preservescalar}
Test structure preserves probabilities, i.e.~$\dax(r)=r$.
\end{lem}
\proof
Consider,
\[\begin{tikzpicture}
	\begin{pgfonlayer}{nodelayer}
		\node [style=point] (0) at (12, -0.4999999) {$\phi$};
		\node [style=none] (1) at (0.9999999, -0) {$=$};
		\node [style=none] (2) at (10, -0) {$=1=$};
		\node [style=copoint] (3) at (12, 0.4999999) {$\phi^\dax$};
		\node [style=point] (4) at (2.5, -0.4999999) {$\psi$};
		\node [style=none] (5) at (0, -0) {$p$};
		\node [style=none] (6) at (5, -0) {where};
		\node [style=copoint] (7) at (2.5, 0.4999999) {$\phi^\dax$};
		\node [style=point] (8) at (8.000001, -0.4999999) {$\psi$};
		\node [style=copoint] (9) at (8.000001, 0.4999999) {$\psi^\dax$};
		\node [style=none] (10) at (13, -0) {};
	\end{pgfonlayer}
	\begin{pgfonlayer}{edgelayer}
		\draw (4) to (7);
		\draw (8) to (9);
		\draw (0) to (3);
	\end{pgfonlayer}
\end{tikzpicture},\]
then  \textsf{tests-produce-probabilities}  implies that $0\leq p\leq 1$. Now consider,
\[\begin{tikzpicture}
	\begin{pgfonlayer}{nodelayer}
		\node [style=copoint] (0) at (4, 0.625) {$\psi^{\dax\ }$};
		\node [style=copoint] (1) at (8.500001, 0.4999999) {$\psi^\dax$};
		\node [style=point] (2) at (4, -0.625) {$\phi^{\dax^\dax}$};
		\node [style=none] (3) at (1.75, -0) {$=$};
		\node [style=none] (4) at (6.499999, -0) {$=$};
		\node [style=point] (5) at (8.500001, -0.4999999) {$\phi$};
		\node [style=none] (6) at (0, -0) {$\dax(p)$};
		\node [style=none] (7) at (9.5, -0) {};
	\end{pgfonlayer}
	\begin{pgfonlayer}{edgelayer}
		\draw (0) to (2);
		\draw (1) to (5);
	\end{pgfonlayer}
\end{tikzpicture},\]
and again   \textsf{tests-produce-probabilities}   implies that $0\leq \dax(p) \leq 1$, and so, $\dax$ preserves $[0,1]$.
Now note that lemmas \ref{lem:InvNormState} and \ref{lem:InvNormEffect} imply that,
\[\begin{tikzpicture}
	\begin{pgfonlayer}{nodelayer}
		\node [style=none] (0) at (1, -0) {$\dax(\dax(p))=$};
		\node [style=none] (1) at (2.75, -0) {$\dax$};
		\node [style=none] (2) at (3.5, -0) {$\dax$};
		\node [style=copoint] (3) at (4.75, 0.4999999) {$\psi^\dax$};
		\node [style=point] (4) at (4.75, -0.4999999) {$\phi$};
		\node [style=none] (5) at (6.75, -0) {$=$};
		\node [style=copoint] (6) at (8, 0.5000001) {$\psi^\dax$};
		\node [style=point] (7) at (8, -0.5000001) {$\phi$};
		\node [style=none] (8) at (9.5, -0) {$=p$};
		\node [style=none] (9) at (4, 1.25) {};
		\node [style=none] (10) at (4, -1.25) {};
		\node [style=none] (11) at (5.5, -1.25) {};
		\node [style=none] (12) at (5.5, 1.25) {};
		\node [style=none] (13) at (3.25, 1.25) {};
		\node [style=none] (14) at (3.25, -1.25) {};
		\node [style=none] (15) at (5.75, 1.25) {};
		\node [style=none] (16) at (5.75, -1.25) {};
		\node [style=none] (17) at (10.25, -0) {};
	\end{pgfonlayer}
	\begin{pgfonlayer}{edgelayer}
		\draw (3) to (4);
		\draw (6) to (7);
		\draw [bend right=15, looseness=1.00] (9.center) to (10.center);
		\draw [bend right=15, looseness=1.00] (13.center) to (14.center);
		\draw [bend left=15, looseness=1.00] (12.center) to (11.center);
		\draw [in=75, out=-75, looseness=1.00] (15.center) to (16.center);
	\end{pgfonlayer}
\end{tikzpicture},\]
so the test structure is involutive for $[0,1]$. Moreover, \textsf{composability} implies that  $\dax$ is multiplicative on all scalars,
\[\dax(r_1r_2)=\dax(r_1)\dax(r_2).\]
The result that $\dax(r)=r$ is an application of standard results regarding functional equations~\cite{FunctionalEquations}. Multiplicativity and preservation of $[0,1]$ imply that $\dax(r)=r^a$ where $a \in \mathds{R}^+$. Involutivity on $[0,1]$ demands that $a=1$ and so $\dax(r)=r$.
\endproof

\begin{thm} For \underline{tomographic} process theories, the test structure is involutive.\label{thm:InvProcess}\end{thm}
\proof
Firstly we extend to all states using \textsf{testability}; write an arbitrary state $\chi = r \psi$ where $r$ is a scalar and $\psi$ normalised. Then lemmas \ref{lem:InvNormState} and \ref{lem:preservescalar} imply that,
\[\chi^{\dax^\dax} = r^{\dax^\dax}\psi^{\dax^\dax}=r\psi = \chi.\]
Tomography then allows us to extend this to all effects, $e$,
\[
\begin{tikzpicture}
	\begin{pgfonlayer}{nodelayer}
		\node [style=copoint] (0) at (1.5, 0.4999999) {$e^{\dax^\dax}$};
		\node [style=point] (1) at (1.5, -0.4999999) {$\chi$};
		\node [style=none] (2) at (-0.4999999, 0.2500001) {$\forall \chi$};
		\node [style=none] (3) at (2.75, -0) {$=$};
		\node [style=none] (4) at (3.5, -0) {$\dax$};
		\node [style=none] (5) at (4.25, -0) {$\dax$};
		\node [style=copoint] (6) at (5.5, 0.4999999) {$e$};
		\node [style=point] (7) at (5.5, -0.4999999) {$\chi$};
		\node [style=none] (8) at (4.75, 1.25) {};
		\node [style=none] (9) at (4.75, -1.25) {};
		\node [style=none] (10) at (4, -1.25) {};
		\node [style=none] (11) at (4, 1.25) {};
		\node [style=none] (12) at (6.25, 1.25) {};
		\node [style=none] (13) at (6.25, -1.25) {};
		\node [style=none] (14) at (6.5, -1.25) {};
		\node [style=none] (15) at (6.5, 1.25) {};
		\node [style=point] (16) at (8.75, -0.4999999) {$\chi$};
		\node [style=none] (17) at (7.5, -0) {$=$};
		\node [style=copoint] (18) at (8.75, 0.4999999) {$e$};
		\node [style=none] (19) at (10.75, -0) {$\iff$};
		\node [style=copoint] (20) at (13, 0.4999999) {$e^{\dax^\dax}$};
		\node [style=none] (21) at (13, -0.4999999) {};
		\node [style=none] (22) at (14.25, -0) {$=$};
		\node [style=copoint] (23) at (15.25, 0.4999999) {$e$};
		\node [style=none] (24) at (15.25, -0.4999999) {};
	\end{pgfonlayer}
	\begin{pgfonlayer}{edgelayer}
		\draw (0) to (1);
		\draw [bend right=15, looseness=1.00] (11.center) to (10.center);
		\draw [in=105, out=-105, looseness=1.00] (8.center) to (9.center);
		\draw [bend left=15, looseness=1.00] (12.center) to (13.center);
		\draw [bend left=15, looseness=1.00] (15.center) to (14.center);
		\draw (6) to (7);
		\draw (18) to (16);
		\draw (20) to (21.center);
		\draw (23) to (24.center);
	\end{pgfonlayer}
\end{tikzpicture}\ \ .
\]
And finally, using tomography again we can extend this to all processes, $f$,
\[\begin{tikzpicture}
	\begin{pgfonlayer}{nodelayer}
		\node [style=none] (0) at (0, 0.5) {$\forall \chi, e, C$};
		\node [style=none] (1) at (5.75, -0) {$=$};
		\node [style=none] (2) at (6.5, -0) {$\dax$};
		\node [style=none] (3) at (7.25, -0) {$\dax$};
		\node [style=none] (4) at (8, 2.25) {};
		\node [style=none] (5) at (8, -2.25) {};
		\node [style=none] (6) at (7.25, -2.25) {};
		\node [style=none] (7) at (7.25, 2.25) {};
		\node [style=none] (8) at (11, 2.25) {};
		\node [style=none] (9) at (11, -2.25) {};
		\node [style=none] (10) at (11.25, -2.25) {};
		\node [style=none] (11) at (11.25, 2.25) {};
		\node [style=none] (12) at (12.75, -0) {$=$};
		\node [style=none] (13) at (18.5, -0) {$\iff$};
		\node [style=none] (14) at (22.75, -0) {$=$};
		\node [style=map] (15) at (2.75, -0) {$f^{\dax^\dax}$};
		\node [style=none] (16) at (2.75, 1) {};
		\node [style=none] (17) at (2, 1) {};
		\node [style=none] (18) at (3.5, 2) {};
		\node [style=none] (19) at (5, 0.9999999) {};
		\node [style=none] (20) at (4.25, 0.9999999) {};
		\node [style=none] (21) at (4.25, -0.9999999) {};
		\node [style=none] (22) at (2.75, -0.9999998) {};
		\node [style=none] (23) at (2, -0.9999998) {};
		\node [style=none] (24) at (3.5, -2) {};
		\node [style=none] (25) at (5, -0.9999999) {};
		\node [style=none] (26) at (3.5, 1.5) {$e$};
		\node [style=none] (27) at (4.75, -0) {$C$};
		\node [style=none] (28) at (3.5, -1.5) {$\chi$};
		\node [style=none] (29) at (5, 0.9999999) {};
		\node [style=map] (30) at (8.75, -0) {$f$};
		\node [style=none] (31) at (8.75, 0.9999999) {};
		\node [style=none] (32) at (10.75, -0) {$C$};
		\node [style=none] (33) at (11, -0.9999999) {};
		\node [style=none] (34) at (8, -0.9999999) {};
		\node [style=none] (35) at (9.5, -2) {};
		\node [style=none] (36) at (9.5, 1.5) {$e$};
		\node [style=none] (37) at (11, 0.9999999) {};
		\node [style=none] (38) at (9.5, -1.5) {$\chi$};
		\node [style=none] (39) at (8, 0.9999999) {};
		\node [style=none] (40) at (9.5, 2) {};
		\node [style=none] (41) at (10.25, 0.9999999) {};
		\node [style=none] (42) at (10.25, -0.9999999) {};
		\node [style=none] (43) at (8.75, -0.9999999) {};
		\node [style=none] (44) at (11, 0.9999999) {};
		\node [style=map] (45) at (14.5, -0) {$f$};
		\node [style=none] (46) at (14.5, 0.9999999) {};
		\node [style=none] (47) at (16.5, -0) {$C$};
		\node [style=none] (48) at (16.75, -0.9999999) {};
		\node [style=none] (49) at (13.75, -0.9999999) {};
		\node [style=none] (50) at (15.25, -2) {};
		\node [style=none] (51) at (15.25, 1.5) {$e$};
		\node [style=none] (52) at (16.75, 0.9999999) {};
		\node [style=none] (53) at (15.25, -1.5) {$\chi$};
		\node [style=none] (54) at (13.75, 0.9999999) {};
		\node [style=none] (55) at (15.25, 2) {};
		\node [style=none] (56) at (16, 0.9999999) {};
		\node [style=none] (57) at (16, -0.9999999) {};
		\node [style=none] (58) at (14.5, -0.9999999) {};
		\node [style=none] (59) at (16.75, 0.9999999) {};
		\node [style=map] (60) at (21, -0) {$f^{\dax^\dax}$};
		\node [style=none] (61) at (21, -0.9999999) {};
		\node [style=none] (62) at (21, 0.9999999) {};
		\node [style=none] (63) at (24, 1) {};
		\node [style=none] (64) at (24, -1) {};
		\node [style=map] (65) at (24, -0) {$f$};
	\end{pgfonlayer}
	\begin{pgfonlayer}{edgelayer}
		\draw [bend right=15, looseness=1.00] (7.center) to (6.center);
		\draw [in=105, out=-105, looseness=1.00] (4.center) to (5.center);
		\draw [bend left=15, looseness=1.00] (8.center) to (9.center);
		\draw [bend left=15, looseness=1.00] (11.center) to (10.center);
		\draw (17.center) to (19.center);
		\draw (19.center) to (18.center);
		\draw (18.center) to (17.center);
		\draw (16.center) to (15);
		\draw (15) to (22.center);
		\draw (23.center) to (25.center);
		\draw (25.center) to (24.center);
		\draw (24.center) to (23.center);
		\draw (20.center) to (21.center);
		\draw (39.center) to (37.center);
		\draw (37.center) to (40.center);
		\draw (40.center) to (39.center);
		\draw (31.center) to (30);
		\draw (30) to (43.center);
		\draw (34.center) to (33.center);
		\draw (33.center) to (35.center);
		\draw (35.center) to (34.center);
		\draw (41.center) to (42.center);
		\draw (54.center) to (52.center);
		\draw (52.center) to (55.center);
		\draw (55.center) to (54.center);
		\draw (46.center) to (45);
		\draw (45) to (58.center);
		\draw (49.center) to (48.center);
		\draw (48.center) to (50.center);
		\draw (50.center) to (49.center);
		\draw (56.center) to (57.center);
		\draw (62.center) to (60);
		\draw (60) to (61.center);
		\draw (63.center) to (65);
		\draw (65) to (64.center);
	\end{pgfonlayer}
\end{tikzpicture}\ \ .
\]
\endproof

\begin{thm} \label{Funct}
For test structures of \underline{tomographic} process theories, the operation $\sharp$ in \textsf{transformability} can always be chosen in a manner such that it is a  dagger.
 \end{thm}
\proof
 Theorem \ref{thm:InvProcess} demonstrates that $\dax$ is involutive, and \textsf{transformability} imposes that if $f:A\to B$ then $f^\dax: B\to A$, therefore for a test structure to provide a dagger all we need to check is that it preserves diagrams. This is easiest to prove by checking that $\dax$ preserves the two  primitive  forms of composition, $\otimes$ and $\circ$.  Consider the action of $\dax$ on the diagram,
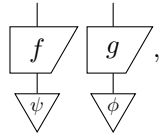
\[\begin{tikzpicture}
	\begin{pgfonlayer}{nodelayer}
		\node [style=point] (0) at (2, -1.5) {$\psi$};
		\node [style=map] (1) at (2, -0) {$f$};
		\node [style=none] (2) at (2, 1.25) {};
		\node [style=point] (3) at (4, -1.5) {$\phi$};
		\node [style=map] (4) at (4, -0) {$g$};
		\node [style=none] (5) at (4, 1.25) {};
	\end{pgfonlayer}
	\begin{pgfonlayer}{edgelayer}
		\draw (0) to (1);
		\draw (1) to (2.center);
		\draw (3) to (4);
		\draw (4) to (5.center);
	\end{pgfonlayer}
\end{tikzpicture},\]
there are two ways to apply the \textsf{composability}  and \textsf{transformability}  axioms to this giving the following constraint,
\begin{equation}\forall \psi, \phi \qquad \begin{tikzpicture}
	\begin{pgfonlayer}{nodelayer}
		\node [style=none] (0) at (-0.7499999, -0) {$\dax$};
		\node [style=point] (1) at (0.9999998, -1.5) {$\psi$};
		\node [style=map] (2) at (0.9999998, -0) {$f$};
		\node [style=none] (3) at (0.9999998, 1.5) {};
		\node [style=point] (4) at (3, -1.5) {$\phi$};
		\node [style=map] (5) at (3, -0) {$g$};
		\node [style=none] (6) at (3, 1.5) {};
		\node [style=none] (7) at (5.25, -0) {$=$};
		\node [style=copoint] (8) at (7, 1.5) {$\psi^\dax$};
		\node [style=copoint] (9) at (9, 1.5) {$\phi^\dax$};
		\node [style=none] (10) at (7, 0.5) {};
		\node [style=none] (11) at (9, 0.5) {};
		\node [style=none] (12) at (10, 0.5) {};
		\node [style=none] (13) at (9.5, -1) {};
		\node [style=none] (14) at (6.5, -1) {};
		\node [style=none] (15) at (6.5, 0.5) {};
		\node [style=none] (16) at (7, -1) {};
		\node [style=none] (17) at (9, -1) {};
		\node [style=none] (18) at (7, -1.75) {};
		\node [style=none] (19) at (9, -1.75) {};
		\node [style=none] (20) at (8, -0.25) {$\dax(f\otimes g)$};
		\node [style=none] (21) at (0.25, 1.75) {};
		\node [style=none] (22) at (0.25, -2.25) {};
		\node [style=none] (23) at (4, -2.25) {};
		\node [style=none] (24) at (4, 1.75) {};
		\node [style=map] (25) at (-7.25, -0.25) {$\dax(f)$};
		\node [style=none] (26) at (-4.25, -1.75) {};
		\node [style=copoint] (27) at (-4.25, 1.25) {$\phi^\dax$};
		\node [style=map] (28) at (-4.25, -0.25) {$\dax(g)$};
		\node [style=none] (29) at (-7.25, -1.75) {};
		\node [style=copoint] (30) at (-7.25, 1.25) {$\psi^\dax$};
		\node [style=none] (31) at (-1.999999, -0) {$=$};
		\node [style=none] (32) at (10.5, -0) {};
	\end{pgfonlayer}
	\begin{pgfonlayer}{edgelayer}
		\draw (1) to (2);
		\draw (2) to (3.center);
		\draw (4) to (5);
		\draw (5) to (6.center);
		\draw (8) to (10.center);
		\draw (9) to (11.center);
		\draw (15.center) to (12.center);
		\draw (12.center) to (13.center);
		\draw (13.center) to (14.center);
		\draw (14.center) to (15.center);
		\draw (16.center) to (18.center);
		\draw (17.center) to (19.center);
		\draw [bend left=15, looseness=1.00] (22.center) to (21.center);
		\draw [bend right=15, looseness=1.00] (23.center) to (24.center);
		\draw (30) to (25);
		\draw (25) to (29.center);
		\draw (27) to (28);
		\draw (28) to (26.center);
	\end{pgfonlayer}
\end{tikzpicture}.\label{C1}\end{equation}
Next consider applying $\dax$ to the diagram,
\[\begin{tikzpicture}
	\begin{pgfonlayer}{nodelayer}
		\node [style=point] (0) at (2, -2.25) {$\psi$};
		\node [style=map] (1) at (2, -0.7499999) {$f$};
		\node [style=map] (2) at (2, 0.7499999) {$g$};
		\node [style=none] (3) at (2, 2.25) {};
		\node [style=none] (4) at (3.5, -0) {};
	\end{pgfonlayer}
	\begin{pgfonlayer}{edgelayer}
		\draw (0) to (1);
		\draw (1) to (2);
		\draw (2) to (3.center);
	\end{pgfonlayer}
\end{tikzpicture},\]
here there are again two different ways to apply the  \textsf{transformability}  axiom, and so we obtain,
\begin{equation} \forall \psi \qquad \begin{tikzpicture}
	\begin{pgfonlayer}{nodelayer}
		\node [style=none] (0) at (0, -0) {$\dax$};
		\node [style=point] (1) at (1.75, -1.75) {$\psi$};
		\node [style=map] (2) at (1.75, -0.25) {$f$};
		\node [style=map] (3) at (1.75, 1.25) {$g$};
		\node [style=none] (4) at (1.75, 2.25) {};
		\node [style=none] (5) at (4.25, -0) {$=$};
		\node [style=copoint] (6) at (7, 1) {$\psi^\dax$};
		\node [style=map] (7) at (7, -0.5) {$\dax(f\circ g)$};
		\node [style=none] (8) at (7, -1.5) {};
		\node [style=map] (9) at (-4, -1.5) {$\dax(g)$};
		\node [style=copoint] (10) at (-4, 1.5) {$\psi^\dax$};
		\node [style=none] (11) at (-4, -2.5) {};
		\node [style=map] (12) at (-4, -0) {$\dax(f)$};
		\node [style=none] (13) at (1, 2.5) {};
		\node [style=none] (14) at (2.75, 2.5) {};
		\node [style=none] (15) at (1, -2.5) {};
		\node [style=none] (16) at (2.75, -2.5) {};
		\node [style=none] (17) at (-1.5, -0) {$=$};
		\node [style=none] (18) at (8.75, -0) {};
		\node [style=none] (19) at (8.75, -0.25) {};
		\node [style=none] (20) at (9.25, -0) {};
	\end{pgfonlayer}
	\begin{pgfonlayer}{edgelayer}
		\draw (1) to (2);
		\draw (2) to (3);
		\draw (3) to (4.center);
		\draw (7) to (6);
		\draw (8.center) to (7);
		\draw (10) to (12);
		\draw (12) to (9);
		\draw (9) to (11.center);
		\draw [bend left=15, looseness=1.00] (15.center) to (13.center);
		\draw [bend right=15, looseness=1.00] (16.center) to (14.center);
	\end{pgfonlayer}
\end{tikzpicture} .\label{C2}\end{equation}
The above two conditions (eq.~\ref{C1} \&~\ref{C2}) are direct consequences of the  test structure  axioms and so must be satisfied for the axioms to hold. There is an obvious solution to these:
\begin{equation}\begin{tikzpicture}
	\begin{pgfonlayer}{nodelayer}
		\node [style=none] (0) at (0, 0.5) {};
		\node [style=none] (1) at (3.25, 0.5) {};
		\node [style=none] (2) at (3, -0.5) {};
		\node [style=none] (3) at (0, -0.5) {};
		\node [style=none] (4) at (0.5, 0.5) {};
		\node [style=none] (5) at (2.5, 0.5) {};
		\node [style=none] (6) at (2.5, -0.5) {};
		\node [style=none] (7) at (0.5, -0.5) {};
		\node [style=none] (8) at (0.4999999, 1.5) {};
		\node [style=none] (9) at (2.5, 1.5) {};
		\node [style=none] (10) at (0.4999999, -1.5) {};
		\node [style=none] (11) at (2.5, -1.5) {};
		\node [style=none] (12) at (1.5, -0) {$\dax(f\otimes g)$};
		\node [style=none] (13) at (4, -0) {$=$};
		\node [style=none] (14) at (6, -1.5) {};
		\node [style=map] (15) at (6, -0) {$\dax(f)$};
		\node [style=none] (16) at (6, 1.5) {};
		\node [style=none] (17) at (9.000001, -1.5) {};
		\node [style=map] (18) at (9, -0) {$\dax(g)$};
		\node [style=none] (19) at (9.000001, 1.5) {};
		\node [style=none] (20) at (12, -0) {\&};
		\node [style=none] (21) at (15, -1.5) {};
		\node [style=map] (22) at (15, -0) {$\dax(g\circ f)$};
		\node [style=none] (23) at (15, 1.5) {};
		\node [style=none] (24) at (18, -0) {$=$};
		\node [style=map] (25) at (20, -0.7499999) {$\dax(g)$};
		\node [style=map] (26) at (20, 0.7499999) {$\dax(f)$};
		\node [style=none] (27) at (20, 2.25) {};
		\node [style=none] (28) at (20, -2.25) {};
		\node [style=none] (29) at (21.75, -0) {};
	\end{pgfonlayer}
	\begin{pgfonlayer}{edgelayer}
		\draw (0.center) to (1.center);
		\draw (1.center) to (2.center);
		\draw (2.center) to (3.center);
		\draw (3.center) to (0.center);
		\draw (4.center) to (8.center);
		\draw (5.center) to (9.center);
		\draw (7.center) to (10.center);
		\draw (6.center) to (11.center);
		\draw (14.center) to (15);
		\draw (15) to (16.center);
		\draw (17.center) to (18);
		\draw (18) to (19.center);
		\draw (21.center) to (22);
		\draw (22) to (23.center);
		\draw (28.center) to (25);
		\draw (25) to (26);
		\draw (26) to (27.center);
	\end{pgfonlayer}
\end{tikzpicture},\label{funct}\end{equation}
and so there will always exist an operation satisfying the axioms as well as eq.~\ref{funct}, and so there is always a $\dax$ which is a process theoretic dagger, $\dax$. \endproof

\begin{thm}
All  test structures of \underline{local tomographic} process theories induce daggers.
\end{thm}
\proof The only solution to equations \ref{C1} \&~\ref{C2} for a tomographically local theory are eq.~\ref{funct} and so $\dax$ must be a dagger.\endproof

In the process-theoretic context, daggers have always been taken to be involutive.   Now, finally, we have shown that there is a good justification for doing so by demonstrating how this important feature of the Hermitian adjoint follows from the notion of a test structure.

 \section{Deriving the Hermitian adjoint for quantum theory\label{QT}}

In this section we demonstrate that for quantum theory any  test structure  must be a Hermitian adjoint.  In order to do so, we will need to extend the process theory representing quantum theory so that it also includes
mixed states, rather than just the pure theory described in the previous sections.

\begin{ex}
We  construct the process theory of \underline{mixed post-selected} quantum processes  from the process theories $\textgoth{D}[\mathds{C}LM]$ and $\mathds{C}LM$. First note that we have an embedding $\textgoth{D}[\mathds{C}LM]\hookrightarrow \mathds{C}LM$.
This embedding allows us to take sums of arbitrary processes of $\textgoth{D}[\mathds{C}LM]$ within $\mathds{C}LM$.
These sums, together with the availability of the scalars as processes, allows one to form all
linear combinations:
\[\begin{tikzpicture}
	\begin{pgfonlayer}{nodelayer}
		\node [style=kpointconj] (0) at (0.7500001, -0.5) {$\psi_i$};
		\node [style=kpoint] (1) at (2.25, -0.5) {$\psi_i$};
		\node [style=none] (2) at (0.7500001, 0.75) {};
		\node [style=none] (3) at (2.25, 0.75) {};
		\node [style=none] (4) at (-2.25, -0) {$\sum_i$};
		\node [style=scalar] (5) at (-1, -0) {$r_i$};
	\end{pgfonlayer}
	\begin{pgfonlayer}{edgelayer}
		\draw (0) to (2.center);
		\draw (1) to (3.center);
	\end{pgfonlayer}
\end{tikzpicture}\ \ .\]
which, in particular, includes all mixtures:
\[\begin{tikzpicture}
	\begin{pgfonlayer}{nodelayer}
		\node [style=kpointconj] (0) at (0.7500001, -0.5) {$\psi_i$};
		\node [style=kpoint] (1) at (2.25, -0.5) {$\psi_i$};
		\node [style=none] (2) at (0.7500001, 0.75) {};
		\node [style=none] (3) at (2.25, 0.75) {};
		\node [style=none] (4) at (-2.25, -0) {$\sum_i$};
		\node [style=scalar] (5) at (-1, -0) {$p_i$};
	\end{pgfonlayer}
	\begin{pgfonlayer}{edgelayer}
		\draw (0) to (2.center);
		\draw (1) to (3.center);
	\end{pgfonlayer}
\end{tikzpicture} \quad \text{where}\quad \sum_ip_i=1\ .\]
We call the resulting process theory $\textgoth{M}[\mathds{C}LM]$. For notational convenience we denote processes in this theory with bold lines, for example:
\[\begin{tikzpicture}
	\begin{pgfonlayer}{nodelayer}
		\node [style=dmap] (0) at (0, -0) {$f$};
		\node [style=none] (1) at (0, 1) {};
		\node [style=none] (2) at (0, -1) {};
		\node [style=none] (3) at (2, -0) {$=$};
		\node [style=none] (4) at (3.25, -0) {$\sum_i$};
		\node [style=scalar] (5) at (4.25, -0) {$r_i$};
		\node [style=mapconj] (6) at (6, -0) {$f_i$};
		\node [style=map] (7) at (7.75, -0) {$f_i$};
		\node [style=none] (8) at (6, 1) {};
		\node [style=none] (9) at (6, -1) {};
		\node [style=none] (10) at (7.75, -1) {};
		\node [style=none] (11) at (7.75, 1) {};
	\end{pgfonlayer}
	\begin{pgfonlayer}{edgelayer}
		\draw [style=doubled] (1.center) to (0);
		\draw [style=doubled] (0) to (2.center);
		\draw (8.center) to (6);
		\draw (6) to (9.center);
		\draw (11.center) to (7);
		\draw (7) to (10.center);
	\end{pgfonlayer}
\end{tikzpicture} \ .\]
\end{ex}

We will now assume that \textsf{transformability} applies to $\textgoth{M}[\mathds{C}LM]$, that is:
\[
\forall f\ \exists f^\dax\quad\text{ s.t. } \quad\left(\begin{tikzpicture}
	\begin{pgfonlayer}{nodelayer}
		\node [style=dpoint] (0) at (0, -1.25) {$\psi$};
		\node [style=dmap] (1) at (0, 0.5) {$f$};
		\node [style=none] (2) at (0, 2) {};
		\node [style=none] (3) at (0.5, -0.5) {$A$};
		\node [style=none] (4) at (0.5, 1.75) {$B$};
	\end{pgfonlayer}
	\begin{pgfonlayer}{edgelayer}
		\draw[style=doubled] (0) to (1);
		\draw [style=doubled](1) to (2.center);
	\end{pgfonlayer}
\end{tikzpicture} \right)^\dax \  =\ \ \ \begin{tikzpicture}
	\begin{pgfonlayer}{nodelayer}
		\node [style=none] (0) at (0, -2.25) {};
		\node [style=dmap] (1) at (0, -0.75) {$f^\dax$};
		\node [style=dcopoint] (2) at (0, 1.25) {$\psi^\dax$};
		\node [style=none] (3) at (0.5, -1.75) {$B$};
		\node [style=none] (4) at (0.5, 0.25) {$A$};
	\end{pgfonlayer}
	\begin{pgfonlayer}{edgelayer}
		\draw [style=doubled](0.center) to (1);
		\draw [style=doubled](1) to (2);
	\end{pgfonlayer}
\end{tikzpicture} \ \ \ .
\]
 where now $f$ can now be an arbitrary sum of processes, $f=\sum_i f_i$. Conceptually, this is a very natural assumption, which states that uncertainty about how a state transforms translates into uncertainty about how the test for that state transforms (see also section~\ref{Section:Additive} below).

Before showing how the test structure provides a Hermitian adjoint we show how another candidate dagger, the transpose, fails to be a test-structure.
\begin{thm} The transpose, $T$,  does not provide a test structure for $\textgoth{M}[\mathds{C}LM]$. \end{thm}
\proof
Consider any states $\psi$ and $\phi$ in $\mathds{C}LM$ such that $\textgoth{D}(\psi)^T\circ \textgoth{D}(\psi)= 1= \textgoth{D}(\phi)^T \circ\textgoth{D}(\phi)$ and, $\textgoth{D}(\psi)^T\circ\textgoth{D}(\phi)=0$ (e.g.~the computational basis states). Then $\textgoth{D}(\psi + i \phi)^T\circ\textgoth{D}(\psi+i \phi) = 1+0+0+i^2= 0$ violating \textsf{testability}.
\endproof

\begin{thm}\label{Theorem:HermAdj}
The test structure provides a Hermitian adjoint for $\textgoth{M}[\mathds{C}LM]$. \end{thm}
\proof
First note that quantum theory is a locally tomographic additive process theory, and so any test structure provides a dagger. This provides an inner product on the states defined as:
\[\langle\psi,\phi\rangle\ :=\ \begin{tikzpicture}
	\begin{pgfonlayer}{nodelayer}
		\node [style=dcopoint] (0) at (0, .5) {$\psi^\dax$};
		\node [style=dpoint] (1) at (0, -.5) {$\phi$};
	\end{pgfonlayer}
	\begin{pgfonlayer}{edgelayer}
		\draw [style=doubled](1) to (0);
	\end{pgfonlayer}
\end{tikzpicture} \ \ .\]
It is simple to check that the inner product axioms are satisfied:\\
Symmetry:
\[\begin{tikzpicture}
	\begin{pgfonlayer}{nodelayer}
		\node [style=none] (0) at (1, -0) {$\left<\psi,\phi\right>=$};
		\node [style=dcopoint] (1) at (3.75, 0.5000001) {$\psi^\dax$};
		\node [style=dpoint] (2) at (3.75, -0.5000001) {$\phi$};
		\node [style=none] (3) at (5.5, -0) {$=$};
		\node [style=none] (4) at (6.5, -0) {$\dax$};
		\node [style=dcopoint] (5) at (8, 0.5000001) {$\psi^\dax$};
		\node [style=dpoint] (6) at (8, -0.5000001) {$\phi$};
		\node [style=none] (7) at (7.25, 1.5) {};
		\node [style=none] (8) at (7.25, -1.25) {};
		\node [style=none] (9) at (8.75, -1.25) {};
		\node [style=none] (10) at (8.75, 1.5) {};
		\node [style=none] (11) at (10, -0) {$=$};
		\node [style=dcopoint] (12) at (11.5, 0.5000001) {$\phi^\dax$};
		\node [style=dpoint] (13) at (11.5, -0.5000001) {$\psi$};
		\node [style=none] (14) at (14, -0) {$=\left<\phi,\psi\right>$};
	\end{pgfonlayer}
	\begin{pgfonlayer}{edgelayer}
		\draw [style=doubled] (2) to (1);
		\draw [style=doubled] (6) to (5);
		\draw [bend left=15, looseness=1.00] (8.center) to (7.center);
		\draw [bend right=15, looseness=1.00] (9.center) to (10.center);
		\draw [style=doubled] (13) to (12);
	\end{pgfonlayer}
\end{tikzpicture}\ \ ,\]
where the second equality follows by lemma \ref{lem:preservescalar} and the third from the extended form of \textsf{transformability}.\\
Linearity:
\[
\begin{tikzpicture}
	\begin{pgfonlayer}{nodelayer}
		\node [style=none] (0) at (1.25, -0) {$\left<\phi,a\psi+b\chi\right>\ =$};
		\node [style=scalar] (1) at (4.75, -0) {$a$};
		\node [style=dcopoint] (2) at (6.25, 0.5000001) {$\phi^\dax$};
		\node [style=dpoint] (3) at (6.25, -0.5000001) {$\psi$};
		\node [style=scalar] (4) at (8.25, -0) {$b$};
		\node [style=dcopoint] (5) at (9.75, 0.5000001) {$\phi^\dax$};
		\node [style=dpoint] (6) at (9.75, -0.5000001) {$\chi$};
		\node [style=none] (7) at (7.25, -0) {$+$};
		\node [style=none] (8) at (14.25, -0) {$=\ a\left<\phi,\psi\right>+b\left<\phi,\chi\right>$};
	\end{pgfonlayer}
	\begin{pgfonlayer}{edgelayer}
		\draw [style=doubled] (3) to (2);
		\draw [style=doubled] (6) to (5);
	\end{pgfonlayer}
\end{tikzpicture}\ \ .
\]
Positivity:
\[
\begin{tikzpicture}
	\begin{pgfonlayer}{nodelayer}
		\node [style=none] (0) at (-6.25, 0) {$\left<\chi,\chi\right>\ =$};
		\node [style=dcopoint] (1) at (-3.5, 0.4999999) {$\chi^\dax$};
		\node [style=dpoint] (2) at (-3.5, -0.4999999) {$\chi$};
		\node [style=none] (3) at (-1.5, 0) {$=$};
		\node [style=dcopoint] (4) at (2, 0.4999999) {$\psi^\dax$};
		\node [style=dpoint] (5) at (2, -0.4999999) {$\psi$};
		\node [style=scalar] (6) at (0.25, 0.5) {$r$};
		\node [style=scalar] (7) at (0.25, -0.5) {$r$};
		\node [style=none] (8) at (4, 0) {$=$};
		\node [style=scalar] (9) at (5.25, -0.5) {$r$};
		\node [style=scalar] (10) at (5.25, 0.5) {$r$};
		\node [style=scalar] (11) at (6, 0) {$1$};
		\node [style=none] (12) at (7.5, 0) {$\geq\  0$};
	\end{pgfonlayer}
	\begin{pgfonlayer}{edgelayer}
		\draw [style=doubled] (2) to (1);
		\draw [style=doubled] (5) to (4);
	\end{pgfonlayer}
\end{tikzpicture}\ \ ,
\]
where the second equation uses  \textsf{testability}, and similarly  positive definiteness follows:
\[
\left<\chi,\chi\right>=0\quad \iff \quad r^2=0 \quad \iff \quad \psi = 0 \ .
\]
The test structure is therefore the Hermitian adjoint associated to this inner product, defined as $\langle\cdot,A\cdot\rangle=\langle A^\dagger\cdot,\cdot\rangle$.
\endproof

This explains why the Hermitian adjoint -- rather than any other dagger such as the transpose -- plays such a prominent role in quantum theory: it has an operational interpretation in terms of a test structure.

\section{Test structures for additive theories}\label{Section:Additive}

Mixed quantum theory $\textgoth{M}[\mathds{C}LM]$ is an example of an additive process theory.

\begin{define} \textbf{ Additive process theories}\label{MPT}
 are process theories that come with a notion of  `sum of diagrams'  that distributes over diagrams:
\[
\begin{tikzpicture}
	\begin{pgfonlayer}{nodelayer}
		\node [style=map] (0) at (-5.25, 0) {$f_i$};
		\node [style=none] (1) at (-5.25, -1.25) {};
		\node [style=none] (2) at (-5.25, 1.25) {};
		\node [style=none] (3) at (-7, 0) {$\sum_i$};
		\node [style=none] (4) at (-7.75, 0.75) {};
		\node [style=none] (5) at (-7.75, -0.75) {};
		\node [style=none] (6) at (-3.75, 0.75) {};
		\node [style=none] (7) at (-3.75, -0.75) {};
		\node [style=none] (8) at (-3, 1.25) {};
		\node [style=none] (9) at (-3, -1.25) {};
		\node [style=none] (10) at (-6.25, 1.25) {};
		\node [style=none] (11) at (-2.25, 1.25) {};
		\node [style=none] (12) at (-1.75, 2.25) {};
		\node [style=none] (13) at (-6.25, 2.25) {};
		\node [style=none] (14) at (-4.25, 2.25) {};
		\node [style=none] (15) at (-4.25, 3) {};
		\node [style=none] (16) at (-2.25, -2.25) {};
		\node [style=none] (17) at (-3, -1.25) {};
		\node [style=none] (18) at (-5.25, -1.25) {};
		\node [style=none] (19) at (-6.25, -2.25) {};
		\node [style=none] (20) at (-3, -1.25) {};
		\node [style=none] (21) at (-6.25, -1.25) {};
		\node [style=none] (22) at (-5.25, -1.25) {};
		\node [style=none] (23) at (-4.25, -2.25) {};
		\node [style=none] (24) at (-1.75, -1.25) {};
		\node [style=none] (25) at (-4.25, -3) {};
		\node [style=none] (26) at (-4.25, -2.25) {};
		\node [style=none] (27) at (5.25, -1.25) {};
		\node [style=none] (28) at (7.5, 1.25) {};
		\node [style=none] (29) at (4.25, 2.25) {};
		\node [style=none] (30) at (5.25, -1.25) {};
		\node [style=none] (31) at (3.75, -3.25) {};
		\node [style=none] (32) at (6.5, -2.25) {};
		\node [style=none] (33) at (7.5, -1.25) {};
		\node [style=none] (34) at (9, 3) {};
		\node [style=none] (35) at (3.75, 3) {};
		\node [style=none] (36) at (5.25, -1.25) {};
		\node [style=none] (37) at (4.25, -2.25) {};
		\node [style=none] (38) at (6.5, 2.25) {};
		\node [style=none] (39) at (6.5, -3) {};
		\node [style=none] (40) at (8.25, -2.25) {};
		\node [style=none] (41) at (7.5, -1.25) {};
		\node [style=none] (42) at (5.25, 1.25) {};
		\node [style=none] (43) at (4.25, -1.25) {};
		\node [style=none] (44) at (8.75, 2.25) {};
		\node [style=none] (45) at (2.25, 0) {$\sum_i$};
		\node [style=none] (46) at (6.5, 3) {};
		\node [style=none] (47) at (4.25, 1.25) {};
		\node [style=map] (48) at (5.25, 0) {$f_i$};
		\node [style=none] (49) at (8.25, 1.25) {};
		\node [style=none] (50) at (7.5, -1.25) {};
		\node [style=none] (51) at (8.75, -1.25) {};
		\node [style=none] (52) at (9, -3.25) {};
		\node [style=none] (53) at (6.5, -2.25) {};
		\node [style=none] (54) at (0, 0) {$=$};
		\node [style=none] (55) at (-4.25, -1.75) {$f$};
		\node [style=none] (56) at (-4.25, 1.75) {$g$};
		\node [style=none] (57) at (6.5, -1.75) {$f$};
		\node [style=none] (58) at (6.5, 1.75) {$g$};
		\node [style=none] (59) at (12.5, 0) {$\forall\ f,\ g,\ C$};
		\node [style=none] (60) at (-2.5, 0) {$C$};
		\node [style=none] (61) at (8, 0) {$C$};
	\end{pgfonlayer}
	\begin{pgfonlayer}{edgelayer}
		\draw (1.center) to (0);
		\draw (0) to (2.center);
		\draw [bend right=15, looseness=1.00] (4.center) to (5.center);
		\draw [bend left=15, looseness=1.00] (6.center) to (7.center);
		\draw (13.center) to (12.center);
		\draw (12.center) to (11.center);
		\draw (11.center) to (10.center);
		\draw (10.center) to (13.center);
		\draw (14.center) to (15.center);
		\draw (8.center) to (9.center);
		\draw (21.center) to (24.center);
		\draw (24.center) to (16.center);
		\draw (16.center) to (19.center);
		\draw (19.center) to (21.center);
		\draw (18.center) to (22.center);
		\draw (20.center) to (17.center);
		\draw (25.center) to (26.center);
		\draw (36.center) to (48);
		\draw (48) to (42.center);
		\draw [bend right=15, looseness=1.00] (35.center) to (31.center);
		\draw [bend left=15, looseness=1.00] (34.center) to (52.center);
		\draw (29.center) to (44.center);
		\draw (44.center) to (49.center);
		\draw (49.center) to (47.center);
		\draw (47.center) to (29.center);
		\draw (46.center) to (38.center);
		\draw (28.center) to (50.center);
		\draw (43.center) to (51.center);
		\draw (51.center) to (40.center);
		\draw (40.center) to (37.center);
		\draw (37.center) to (43.center);
		\draw (27.center) to (30.center);
		\draw (41.center) to (33.center);
		\draw (32.center) to (39.center);
	\end{pgfonlayer}
\end{tikzpicture}.
\]
\end{define}

\begin{rem}  \em
In category-theoretic terms, additivity of process theories means enrichment in commutative monoids \cite{coecke2014picturing}.  That the numbers are positive reals means that the morphisms from the tensor unit to itself are the positive reals.
\end{rem}

Having both sums and numbers, together with the fact that since numbers have no outputs they can freely move around in diagrams, it also follows that:

\begin{prop}
In additive process theories  convex combinations (or equivalent, probabilistic mixtures) distribute over diagrams:
\[
\begin{tikzpicture}
	\begin{pgfonlayer}{nodelayer}
		\node [style=map] (0) at (-5.25, 0) {$f_i$};
		\node [style=none] (1) at (-5.25, -1.25) {};
		\node [style=none] (2) at (-5.25, 1.25) {};
		\node [style=none] (3) at (-7.5, 0) {$\sum_i p_i$};
		\node [style=none] (4) at (-8.5, 0.75) {};
		\node [style=none] (5) at (-8.5, -0.75) {};
		\node [style=none] (6) at (-3.75, 0.75) {};
		\node [style=none] (7) at (-3.75, -0.75) {};
		\node [style=none] (8) at (-3, 1.25) {};
		\node [style=none] (9) at (-3, -1.25) {};
		\node [style=none] (10) at (-6.25, 1.25) {};
		\node [style=none] (11) at (-2, 1.25) {};
		\node [style=none] (12) at (-2, 2.25) {};
		\node [style=none] (13) at (-6.25, 2.25) {};
		\node [style=none] (14) at (-4.25, 2.25) {};
		\node [style=none] (15) at (-4.25, 3) {};
		\node [style=none] (16) at (-2, -2.25) {};
		\node [style=none] (17) at (-3, -1.25) {};
		\node [style=none] (18) at (-5.25, -1.25) {};
		\node [style=none] (19) at (-6.25, -2.25) {};
		\node [style=none] (20) at (-3, -1.25) {};
		\node [style=none] (21) at (-6.25, -1.25) {};
		\node [style=none] (22) at (-5.25, -1.25) {};
		\node [style=none] (23) at (-4.25, -2.25) {};
		\node [style=none] (24) at (-2, -1.25) {};
		\node [style=none] (25) at (-4.25, -3) {};
		\node [style=none] (26) at (-4.25, -2.25) {};
		\node [style=none] (27) at (5.75, -1.25) {};
		\node [style=none] (28) at (8, 1.25) {};
		\node [style=none] (29) at (4.75, 2.25) {};
		\node [style=none] (30) at (5.75, -1.25) {};
		\node [style=none] (31) at (4.25, -3.25) {};
		\node [style=none] (32) at (7, -2.25) {};
		\node [style=none] (33) at (8, -1.25) {};
		\node [style=none] (34) at (9.5, 3) {};
		\node [style=none] (35) at (4.25, 3) {};
		\node [style=none] (36) at (5.75, -1.25) {};
		\node [style=none] (37) at (4.75, -2.25) {};
		\node [style=none] (38) at (7, 2.25) {};
		\node [style=none] (39) at (7, -3) {};
		\node [style=none] (40) at (9, -2.25) {};
		\node [style=none] (41) at (8, -1.25) {};
		\node [style=none] (42) at (5.75, 1.25) {};
		\node [style=none] (43) at (4.75, -1.25) {};
		\node [style=none] (44) at (9, 2.25) {};
		\node [style=none] (45) at (2.25, 0) {$\sum_i p_i$};
		\node [style=none] (46) at (7, 3) {};
		\node [style=none] (47) at (4.75, 1.25) {};
		\node [style=map] (48) at (5.75, 0) {$f_i$};
		\node [style=none] (49) at (9, 1.25) {};
		\node [style=none] (50) at (8, -1.25) {};
		\node [style=none] (51) at (9, -1.25) {};
		\node [style=none] (52) at (9.5, -3.25) {};
		\node [style=none] (53) at (7, -2.25) {};
		\node [style=none] (54) at (0, 0) {$=$};
		\node [style=none] (55) at (-4.25, -1.75) {$f$};
		\node [style=none] (56) at (-4.25, 1.75) {$g$};
		\node [style=none] (57) at (7, -1.75) {$f$};
		\node [style=none] (58) at (7, 1.75) {$g$};
		\node [style=none] (59) at (12.5, 0) {$\forall\ f,\ g,\ C$};
		\node [style=none] (60) at (-2.5, 0) {$C$};
		\node [style=none] (61) at (8.5, 0) {$C$};
	\end{pgfonlayer}
	\begin{pgfonlayer}{edgelayer}
		\draw (1.center) to (0);
		\draw (0) to (2.center);
		\draw [bend right=15, looseness=1.00] (4.center) to (5.center);
		\draw [bend left=15, looseness=1.00] (6.center) to (7.center);
		\draw (13.center) to (12.center);
		\draw (12.center) to (11.center);
		\draw (11.center) to (10.center);
		\draw (10.center) to (13.center);
		\draw (14.center) to (15.center);
		\draw (8.center) to (9.center);
		\draw (21.center) to (24.center);
		\draw (24.center) to (16.center);
		\draw (16.center) to (19.center);
		\draw (19.center) to (21.center); 
		\draw (18.center) to (22.center);
		\draw (20.center) to (17.center);
		\draw (25.center) to (26.center);
		\draw (36.center) to (48);
		\draw (48) to (42.center);
		\draw [bend right=15, looseness=1.00] (35.center) to (31.center);
		\draw [bend left=15, looseness=1.00] (34.center) to (52.center);
		\draw (29.center) to (44.center);
		\draw (44.center) to (49.center);
		\draw (49.center) to (47.center);
		\draw (47.center) to (29.center);
		\draw (46.center) to (38.center);
		\draw (28.center) to (50.center);
		\draw (43.center) to (51.center);
		\draw (51.center) to (40.center);
		\draw (40.center) to (37.center);
		\draw (37.center) to (43.center);
		\draw (27.center) to (30.center);
		\draw (41.center) to (33.center);
		\draw (32.center) to (39.center);
	\end{pgfonlayer}
\end{tikzpicture},
\]
where $\sum_ip_i=1$.
\end{prop}

The notion of an  additive  process theory is similar to the well-studied framework of generalised probabilistic theories \cite{Barrett, barnum2011information}. There are many presentations of this framework which are subtly different  but  in their most recent incarnation they have the structure of an additive process theory  \cite{chiribella2011informational}. While the usual GPTs have  no unique parallel composition operation,  the underlying structure of the processes allows one nonetheless to talk about composite systems.

\begin{thm}
Additive process theories do not have test structures.
\end{thm}
\proof
Consider a state that is a convex mixture of normalised states and assume the existence of a test structure,
\[\begin{tikzpicture}
	\begin{pgfonlayer}{nodelayer}
		\node [style=point] (0) at (-0.4999999, -0.2500001) {$\psi$};
		\node [style=none] (1) at (-0.4999999, 0.9999999) {};
		\node [style=none] (2) at (1, -0) {$=$};
		\node [style=none] (3) at (2.5, -0) {$\sum_i p_i$};
		\node [style=point] (4) at (4.25, -0.2500001) {$\psi_i$};
		\node [style=none] (5) at (4.25, 0.9999999) {};
	\end{pgfonlayer}
	\begin{pgfonlayer}{edgelayer}
		\draw (1.center) to (0);
		\draw (5.center) to (4);
	\end{pgfonlayer}
\end{tikzpicture}\ \ ,\]
where $\sum_ip_i=1$. Then if we demand that we have a test for that state, $\psi^\dax$, then by definition this test must produce probabilities,
\[\begin{tikzpicture}
	\begin{pgfonlayer}{nodelayer}
		\node [style=point] (0) at (-0.25, -0) {$\psi$};
		\node [style=copoint] (1) at (-0.25, 0.9999999) {$e$};
		\node [style=none] (2) at (8.5, 0.4999999) {$=$};
		\node [style=none] (3) at (5.5, 0.5) {$\sum_i p_i$};
		\node [style=point] (4) at (7.25, -0) {$\psi_i$};
		\node [style=copoint] (5) at (7.25, 0.9999999) {$e$};
		\node [style=none] (6) at (1, 0.5) {$=$};
		\node [style=none] (7) at (2, 0.5) {$1$};
		\node [style=none] (8) at (3.5, 0.5) {$\implies$};
		\node [style=none] (9) at (9.5, 0.4999999) {$1$};
		\node [style=point] (10) at (5.5, -2.5) {$\psi_i$};
		\node [style=none] (11) at (3.5, -2) {$\implies$};
		\node [style=none] (12) at (6.75, -2) {$=$};
		\node [style=copoint] (13) at (5.5, -1.5) {$e$};
		\node [style=none] (14) at (7.75, -2) {$1\ ,$};
	\end{pgfonlayer}
	\begin{pgfonlayer}{edgelayer}
		\draw (1) to (0);
		\draw (5) to (4);
		\draw (13) to (10);
	\end{pgfonlayer}
\end{tikzpicture} \]
which violates sharpness giving a contradiction.
\endproof

This is not surprising; one cannot deterministically test for a probabilistic mixture of states.
However in the previous section we showed that this wasn't a problem for quantum theory, similarly, theorem~\ref{Theorem:HermAdj} can be extended to additive process theories.

\begin{thm}
A process theory with a test structure, embedded within an additive theory, has an inner product defined as:
\[\langle\psi,\phi\rangle:=\begin{tikzpicture}
	\begin{pgfonlayer}{nodelayer}
		\node [style=copoint] (0) at (0, .5) {$\psi^\dax$};
		\node [style=point] (1) at (0, -.5) {$\phi$};
	\end{pgfonlayer}
	\begin{pgfonlayer}{edgelayer}
		\draw [style=none](1) to (0);
	\end{pgfonlayer}
\end{tikzpicture} \ \ ,\]
if we demand that \textsf{transformability} extends to all processes in the additive theory.
\end{thm}
\proof Identical to the quantum case in theorem~\ref{Theorem:HermAdj}. \endproof

\section{Summary and outlook}

In this paper we considered the physical principle of:
\begin{center}
\emph{\textbf{ for  each state there exists a  corresponding `test'.}}
\end{center}

We show that for tomographic process theories this principle leads to a `test structure' which we show provides a process-theoretic dagger. This explains why the dagger should have various properties, involutivity being a particularly surprising consequence given the  test structure's definition. Moreover, for pure quantum theory, we show that the particular dagger provided is the Hermitian adjoint, explaining why this plays such an important role both  in the process-theoretic description of quantum theory,  and of course, in quantum theory itself in the form of the Hilbert space inner-product.

In the final section we begin to explore the connections between test structures  and  purity of a theory, showing that theories with mixed states cannot have a test structure. In fact, if we  consider process theories that come with a discarding operation -- for which we require causality \cite{chiribella2010probabilistic, coecke2014terminality} --
 then we can turn this around and show that a test structure characterises when a process is pure by its testability.

\end{document}